\documentclass[preprint,12pt,authoryear]{elsarticle}

\usepackage[framemethod=tikz]{mdframed}
\usepackage{ntheorem}

\theorembodyfont{\upshape}
\newmdtheoremenv[
	innerleftmargin=10pt,
	innerrightmargin=10pt,
	innertopmargin=10pt,
]{customExample}{Example}

\usepackage{graphicx}
\usepackage[english]{babel}
\usepackage[utf8]{inputenc}
\usepackage[hyphens]{url}
\usepackage{float}
\usepackage{subfig}
\usepackage{enumitem}
\usepackage{listings} 

\usepackage{xcolor}
\usepackage{graphicx}
\usepackage[left=1in,right=1in]{geometry}

\definecolor{sdisagree}{RGB}{195, 35, 20}
\definecolor{disagree}{RGB}{230, 170, 160}
\definecolor{neutral}{RGB}{225, 225, 225}
\definecolor{agree}{RGB}{120, 175, 230}
\definecolor{sagree}{RGB}{35, 110, 195}

\newcommand{\likertpct}[6][0.5]{%
	\begin{tabular}{rcr}
		\the\numexpr#2+#3\relax &
		\resizebox{#1\textwidth}{\height}{%
			\color{sdisagree}\rule{#2mm}{10pt}\color{disagree}\rule{#3mm}{10pt}\color{neutral}\rule{#4mm}{10pt}\color{agree}\rule{#5mm}{10pt}\color{sagree}\rule{#6mm}{10pt}%
		} &
		\the\numexpr#5+#6\relax\\
	\end{tabular}%
}

\usepackage{array}
\usepackage[flushleft]{threeparttable}
\usepackage{pgf-pie}  
\usepackage{pgfplots}
\def\checkmark{\tikz\fill[scale=0.4](0,.35) -- (.25,0) -- (1,.7) -- (.25,.15) -- cycle;}

\begin{document}
	
\begin{frontmatter}
\title{Flexible Process Variant Binding in Information Systems with Software Product Line Engineering}

\author[1]{Philipp Hehnle\corref{cor1}}
\ead{philipp.hehnle@uni-ulm.de}

\author[1]{Manfred Reichert}
\ead{manfred.reichert@uni-ulm.de}

\cortext[cor1]{Corresponding author}

\affiliation[1]{organization={Institute of Databases and Information Systems, Ulm University},
	addressline={James-Franck-Ring},
	postcode={89081},
	city={Ulm},
	country={Germany}}

\begin{abstract}
Different organisations often run similar digitised business processes to achieve their business goals.
However, organisations often need to slightly adapt the business processes implemented in an information system in order to adopt them.
Various approaches have been proposed to manage variants in process models.
While these approaches mainly deal with control flow variability, in previous work we introduced an approach to manage implementation variants of digitised business processes.
In this context Software Product Line (SPL) Engineering was applied to manage a set of common core artefacts including a process model from which Process-Aware Information Systems (PAIS) can be derived, which differ in the implementation of their process activities.
When deriving a PAIS, implementations are selected for each process activity and then included in the PAIS at compilation time.
One challenge that has not yet been solved is giving users of digitised business processes the option of selecting multiple implementations at runtime.
This paper extends our previous work by not only allowing for the selection of activity implementations at compile time, but also at start time and runtime.
Consequently, it becomes possible to defer the decision as to which implementation should be selected to start time and runtime.
Furthermore, multiple implementations of a particular activity may be selected and executed concurrently.
The presented approach also allows customising the input and output data of activities.
Data from expert interviews with German municipalities suggests digitising business processes with varying implementations is a widespread challenge and our approach is a way to mitigate it.
\end{abstract}


\end{frontmatter}

\section{Introduction}
The right to self-administration of German municipalities leads to various variants of the same business process.
For example, craftspersons may apply for a special parking permit, which allows them to park in areas in which regular citizen have to pay or in which generally parking is not permitted.
Special parking permits for craftspersons are available, inter alia, in the German cities of Munich\footnote{\url{http://www.muenchen.de/dienstleistungsfinder/muenchen/1072021/}}, Constance\footnote{\url{https://www.konstanz.de/serviceportal/-/leistungen+von+a-z/handwerkerparkausweis-beantragen/vbid6000803}}, and Stuttgart\footnote{\url{https://www.stuttgart.de/vv/leistungen/sonderparkausweise-fuer-gewerbetreibende-und-soziale-dienste.php}}.
The business process for checking an application for the special parking permit is similar between the municipalities.
However, there are slight differences among the municipalities.
In some municipalities, the special parking permit is issued automatically, whereas in others a municipal employee needs to issue the parking permit manually.
Other researchers have observed and investigated business process variability in municipalities as well \citep{LaROSA.2017}, focusing on control flow variability.

Approaches have been proposed to deal with business process variability.
Reference processes \citep{vanderAalst.2006} and base processes \citep{AlenaHallerbach.2008}, respectively, were introduced that may be configured to meet the needs of an individual organisation.
However, these approaches focus on variations in the control flow of a business process rather than on the implementation level.
Software Product Line (SPL) Engineering \citep{Bass.2013} addresses the challenge of developing and maintaining a set of similar software products.
A Software Product Line comprises a common set of core artefacts from which features may be selected in order to build a specific software product.
In literature, the process of selecting features from an SPL and building a software product is referred to as \textit{product derivation} \citep{Deelstra.2005,Kastner.2013}.
A Process-Aware Information System (PAIS) corresponds to a software product that executes a business process involving human actors, applications, and information sources \citep{Dumas.2005}.
In previous work \citep{Hehnle.2023}, we applied and simplified the concepts of SPL Engineering to PAISs (\textit{PAIS Product Line}).
At build time, any implementation can be selected for an activity.
However, this static approach lacks flexibility, which is of utmost importance in practice.

\subsection{Problem Statement}
In our previous approach, the activities of a PAIS Product Line have specified input and output data structures, i.e. it is specified what data an activity expects and what data can be collected during the execution of an activity.
The combined input and output data structures of all activities are denoted as \textit{process data structure}.
The process data structure of the PAIS Product Line is assumed immutable, i.e. when deriving a product from a PAIS Product Line the process data structure cannot be customised.
Data structures are implemented in source code (e.g. Java Classes).
If the process data structure is to be customised during product derivation (e.g. a user form is adapted in order to collect different data), the corresponding source code (Java Classes) needs to be adapted, e.g. removing or adding fields in a Java Class.
This poses a challenge as removing fields of a Java Class might lead to compile errors when these fields are used elsewhere.

Furthermore, the implementations for the activities need to be selected at build time.
Selection at start time or runtime is not possible.
In order to give the process engineer of a business process the option of selecting features, the decision as to which activity implementations are selected must be deferred to startup time or runtime.
In previous work, we used the build tool \textit{Apache Maven}\footnote{\url{https://maven.apache.org/}} to conditionally package the selected activity implementations.
When activity implementations are to be selected at startup time or runtime a build tool cannot be used anymore.
The challenge is to find and adopt a mechanism from SPL Engineering allowing for the selection of activity implementations at startup time and runtime.
However, note that when it becomes possible to select and, consequently, deselect activity implementations at runtime, it needs to be specified what shall happen with the data of a running implementation that gets deselected.

Finally, so far only one implementation may be selected for an activity.
In order to give the process engineer of a business process the option of selecting multiple implementations for one activity, it becomes necessary to execute the selected implementations concurrently.
Concurrent running activity implementations might access the same data which can result in unintended overwriting of data.

\subsection{Contribution}
This work addresses the aforementioned limitations and problems of PAIS Product Lines.
Building on previous work \citep{Hehnle.2023}, the main contribution of this work is threefold:

\begin{enumerate}
	\item Concepts and tools from SPL Engineering are adopted and applied to the approach of PAIS Product Lines to enable customising the process data structure when deriving a PAIS product.
	\item Techniques known from SPL Engineering are selected and combined to enable selecting activity implementations at build time, startup time, and runtime.
	\item An approach is presented that allows selecting and running multiple activity implementations concurrently, while ensuring that no data is unintentionally overwritten.
\end{enumerate}

Moreover, it is assured that the process of designing, developing, selecting activity implementations, and deriving products (known as the phases of SPL Engineering) is tool supported.
An evaluation in the form of expert interviews with German municipalities indicates that our contributions tackle a ubiquitous problem and can be applied to a variety of municipal business processes in Germany.

\subsection{Outline}
The remainder of this paper is structured as follows:
Section \ref{sec:fundamentals} presents basic concepts of SPL Engineering, which lay the foundation of the approach, and which are necessary for understanding it.
The approach of PAIS Product Lines we introduced in \citep{Hehnle.2023} is described in Section \ref{sec:background}.
In Section \ref{sec:approach}, this approach is extended to enable more flexibility in that the data structure may be customised during PAIS Product derivation, multiple implementations for one activity may be selected, and the implementation may be bound at different times.
The flexibility of the presented approach is discussed in Section \ref{sec:discussion}.
We evaluate our approach by conducting expert interviews with German municipalities in terms of whether the identified problem is widespread and whether our approach mitigates it.
Then, we evaluate the validity of our approach.
The results are presented in Section \ref{sect:evaluation}.
Section \ref{sec:related-work} discusses related work.
Section \ref{sec:conclusion} concludes the paper and provides and outlook on open challenges and further research topics.
\section{Fundamentals}
\label{sec:fundamentals}

In this section, basic concepts of SPL Engineering are presented, which constitute the foundation of our approach.
The concepts deal with variability among similar software products that may be derived from a set of common core artefacts.
First, the development process (i.e. phases) of an SPL is outlined.
Then, according to the phases, it is presented how to collect and sketch the requirements of an SPL as models.
Features that represent these requirements can be selected to be included in individual-derived software products, which is called feature binding.
Therefore, different feature binding approaches are introduced on an abstract level, before presenting two concrete techniques for feature binding, which are called \textit{variability mechanisms}.
\textit{Feature interactions} are unintended behaviours which pose a problem when binding different features and are discussed followed by presenting tool support for the entire development process of an SPL (i.e. SPL phases).

\subsection{Phases of Software Product Lines Engineering}
\label{subsect:phases}
As opposed to single software products, when developing an SPL, the requirements of various similar, yet different, software products need to be collected, managed and implemented.
SPL Engineering can be divided into four phases \citep{Thum.2014,Kastner.2013}:

\begin{enumerate}
	\item \textit{Domain analysis} includes the activities of collecting the requirements and describing the domain features of the SPL, i.e. collecting the requirements for software products that can be derived from the SPL.
	\item \textit{Domain implementation} deals with implementing the features described during domain analysis in a way that allows composing them individually for each derived software product. 
	\item \textit{Requirements analysis} deals with selecting the features from the domain analysis for a specific software product derived from the SPL. The selected features form a configuration.
	\item \textit{Software generation} deals with building a software product by composing those features specified in the configuration from the requirements analysis.
\end{enumerate}

\subsection{Variability Modelling}
\label{sect:var-model}

Kang et al. \citep{Kang.1990} introduced a notation to capture the functionality of a software product from the perspective of a user.
By applying this notation, it becomes possible to outline the features of a software product as well as their relationship in a tree-structured \textit{feature model}.
The root node represents the software product.
Features of the software product are connected by edges to the root node.
Features themselves may consists of other features and are inter-connected with them via edges as well.
Little circles at the end of an edge refer to optional features, whereas
alternative features are identified by arcs between their corresponding edges.
Feature diagrams help developers to identify what has to be configured in a software product.

\begin{figure}
	\centering
	\setlength{\tabcolsep}{0pt}
	\renewcommand{\arraystretch}{0}
	\begin{tabular}[b]{p{4.1cm}}
		\centering
		\includegraphics[width=32mm]{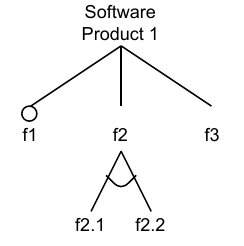} \\
		\small (a) Original Notation \citep{Kang.1990}
	\end{tabular}
	\begin{tabular}[b]{p{4.8cm}}
		\centering
		\includegraphics[width=39mm]{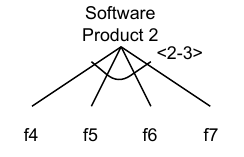} \\
		\small (b) Cardinalities \citep{Czarnecki.2004}
	\end{tabular}
	\begin{tabular}[b]{p{4.5cm}}
		\centering
		\includegraphics[width=42mm]{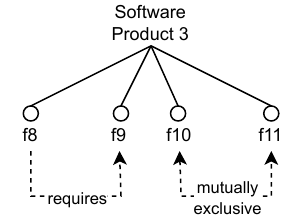} \\
		\small (c)\textit{requires} and \textit{exclusive} Constraint
	\end{tabular}
	\begin{tabular}[t]{p{2.7cm}}
		\centering
		\includegraphics[width=20mm]{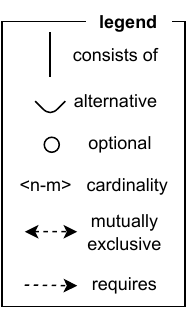} \\
	\end{tabular}
	\setlength{\tabcolsep}{6pt}
	\renewcommand{\arraystretch}{1}
	\caption{Feature Model Notation}
	\label{fig:foda}
\end{figure}

An example of a feature model is displayed in Figure \ref{fig:foda}a.
\textit{Software Product 1} consists of three features, i.e. Features f1, f2, and f3.
While f1 is optional, f2 and f3 are mandatory.
Furthermore, f2 consists of either Feature f2.1 or f2.2.

Besides optional and mandatory features, groups of alternative features may have assigned cardinalities \citep{Czarnecki.2004,Czarnecki.2005} (or multiplicities \citep{Riebisch.}, respectively).
The cardinality of a feature group defines the number of features to be selected from this group.
The feature model depicted in Figure \ref{fig:foda}b represents \textit{Software Product 2}, which comprises four alternative features (i.e. f4, f5, f6, and f7).
The cardinality \textit{\textlangle2-3\textrangle} specifies that for \textit{Software Product 2} at least two features and at most three features have to be selected.

In \citep{Kang.1990}, typical constraints between features such as \textit{requires} or \textit{mutually exclusive} are described in text form.
It is now common practice to use a dashed arrow to indicate that a feature requires another one and a dashed double-headed arrow to illustrate that the two marked features must not be co-selected (e.g. \cite{Riebisch.}, \cite{Benavides.2010}, and \cite{Schobbens.2007}).
Finally, Figure \ref{fig:foda}c shows a feature model that contains a software product with three optional features (i.e. f8, f9, f10, f11).
Due to the constraint \textit{requires}, when selecting Feature f8, Feature f9 must be selected as well.
Furthermore, Features f10 and f11 cannot be co-selected due to the constraint \textit{mutually exclusive}.

\subsection{Feature Binding}
\label{subsect:binding}
In order to use an optional feature of an SPL, it has to be selected in respective software product.
This is called \textit{feature binding} \citep{Czarnecki.2000}.
Features may be bound at different times.
In this context, \citep{Kang.1990} distinguishes three feature binding times: compile time, start time, and runtime. Features bound at compile time are composed during source code compilation, whereas features bound at start time are selected when a software product is launched. Finally, features bound at runtime may be exchanged while the software product is running. Propositions such as \citep{Rosenmuller.2008} and \citep{Rosenmuller.2011b} categorise feature binding times into static binding (i.e. feature binding before software product execution) and dynamic binding (i.e. feature binding during start or while running a software product).
Static and dynamic binding each have advantages and disadvantages \citep{Bosch.2002}:
On the one hand, dynamic binding entails flexibility with respect to adding and removing features at runtime.
On the other hand, this also implies that the software product contains all features.
The size of a software product has an impact on the required resources such as memory and CPU.
In contrast, static binding requires fewer resources but lacks flexibility at runtime.  

Dynamic binding implies adding and removing features during runtime.
Hence, it becomes necessary to specify what happens if a running feature shall be removed.
\citep{Lee.2006} summarises the options:
\begin{itemize}
	\item The running feature is removed immediately.
	\item The current state of the feature is saved before its removal.
	\item The running feature is removed after its execution has completed.
	\item The running feature is not removed at all.
\end{itemize}

Furthermore, the removal of a running feature might impact other features with which it shares data or system processes (i.e. background tasks of the operating system) \citep{Lee.2006}.
\cite{Lee.2006} propose grouping features into feature binding units at development time that can be bound and unbound at runtime.
The feature binding units consider implications of data and system processes the features to be unbound are working on.

A \textit{variation point} corresponds to the location in an information system at which variability occurs, i.e. features are bound.
Methods to implement a variation point (i.e. binding a feature) are called \textit{variability mechanisms} \citep{Jacobson.1997}.
In earlier approaches, the binding time of a feature had to be selected at design time by implementing the variability with an appropriate variability mechanism \citep{Chakravarthy.2008,Rosenmuller.2008,Bosch.2002}.

The approach presented in \citep{Rosenmuller.2008} aims to support static and dynamic feature binding without deciding at design time.
The code basis of the SPL is developed independently of the feature binding time by using feature refinements.
If the features of the SPL shall be bound statically, existing tools are used for composing the feature refinements.
In order to bind the features dynamically, the code base is transformed to use the decorator pattern (also known as wrapper \citep{Gamma.1995}).
Features can be selected after compile time by using factory methods of the feature decorators.
Thus, all features are included at compile time and available afterwards. 

The approach allows selecting either static or dynamic feature binding for all features.
In contrast, \cite{Rosenmuller.2011b} proposes an approach that builds on the results of \cite{Rosenmuller.2008} but allows selecting the binding time per feature.
Changing the order in which features are bound may change their behaviour.
Consider Example \ref{xmpl:feature-binding}.

\begin{customExample} %
	\label{xmpl:feature-binding}
	(Feature Binding Order): Assume Feature A needs to be executed before Feature B.
	Then, Feature A needs to be bound after Feature B.
	However, if Feature A is bound statically and Feature B dynamically the order is reversed.
	The approach presented in \citep{Rosenmuller.2011b} ensures the execution order of the features even when features have different binding times.
	This is achieved by generating hook methods that can be overridden.
\end{customExample}

In dynamic feature binding, mutually exclusive features are included together in the software product as they are not bound before runtime.
Therefore, when binding features at runtime, given constraints (e.g. mutually exclusion and implication) need to be obeyed.
Both \cite{Rosenmuller.2008} and \cite{Rosenmuller.2011b} use a runtime API that enables checking against the corresponding feature model on whether or not the selected features in an SPL Product are consistent.

\textit{Edicts} constitute aspects of aspect-oriented programming \citep{Chakravarthy.2008}.
For a variation point there may be various edicts.
If a feature shall be bound at compile time an edict containing this feature is included.
In contrast, if the implementation of the variation point shall be determined at runtime, an edict containing the code of a design pattern (e.g. decorator pattern) that enables choosing a feature at runtime is included.

\subsection{FeatureHouse}
\label{subsec:featurehouse}
FeatureHouse \citep{Apel.2009b} is variability mechanism for static feature binding and a framework and tool chain for composing software artefacts to software systems.
It uses superimposition to merge the software artefacts, i.e.
the substructures of the software artefacts are merged in order to compose the software system.
The hierarchical structure of a software artefact is represented by a feature structure tree (FST).
For example, a Java artefact may consist of packages, classes, and methods that correspond to FST nodes.
Nodes may be non-terminals (i.e. having child nodes) or terminals.
Depending on the hierarchy level at which the artefacts shall be merged, the structural elements of the latter need to be chosen as terminal nodes.
If, for example, Java artefacts shall be merged at the class level, the classes may be the terminal nodes.
Superimposing two FSTs entails merging their nodes beginning with the root node.
FeatureHouse supports the composition of artefacts written in various languages,
including Java, C\#, and XML.

\subsection{Feature Refinement}
\label{subsect:featurerefinement}
A \textit{feature refinement} \citep{Batory.2004} corresponds to a program increment that represents a feature within related software products, e.g. a software product line.
Feature refinements comprise software fragments that can be incrementally composed to build a software product whereby the software fragments may be source code and other artefacts.
Jak \citep{Batory.2004} is an extension of the programming language Java that contains key words for feature refinement.
A class contained in a Jak file may be refined by another Jak file using the modifier ``refines'', e.g. class members and methods may be added.

The AHEAD tools presented in \citep{Batory.2004}, in turn, are able to compose and translate Jak files to Java files, which constitutes another variability mechanism for static feature binding.
The AHEAD tools also support incrementally composing XML documents written in XAK \citep{Anfurrutia.2007}. 
XAK is a language to refine XML documents and provides the attribute "xak:module" to mark a tag as a module (i.e. this tag may be refined) in the base XML file, which is the starting point for composing XML documents.
All tags in the XML tree below the module are considered implementation and must not be refined.
In separate XAK files, refinements of a module (i.e. XML increment) may be specified.
During composition, the refinements are applied to their corresponding module tag in the base XML file, i.e. the tags in the XML-refinement are appended to the corresponding module tag.
As XAK refinement files constitute only an increment rather than a complete XML, the files are mostly not schema-compliant. 

\subsection{Feature Interactions}
\label{sect:feature-interaction}
In a feature interaction \citep{BowenT.F..1989,Cameron.1993}, in combination multiple features might behave differently than in isolation.
Feature interactions have been subject to research in many domains for a long time \citep{BowenT.F..1989,Cameron.1993,Calder.2003}.
An example of a feature interaction is given by \citep{BowenT.F..1989} and described in Example \ref{xmpl:feature-interaction}:

\begin{customExample} 
	\label{xmpl:feature-interaction}
	(Feature Interaction in Telecommunication): Consider the features call-forwarding and call-waiting on a busy phone line.
	If a phone consists of one of the two features, it works fine.
	However, the behaviour of a phone comprising both features is unclear.
	Either way the requirement of one of the features is not satisfied, maybe even the requirements of both.
\end{customExample}

Feature interactions pose a challenge in SPL Engineering as well \citep{Oster.2011,Cohen.2008,Sahid.2016}.
As opposed to regular Software Engineering, in SPL Engineering, features are selected and composed during product derivation, i.e. not all features will be present in every derived software product.
Consequently, feature interaction might occur only in some of the derived software products.
Testing all derivable software products, however, is not feasible from a practical perspective due to the exponential growth of derivable software products with increasing number of features.
Instead of testing all derivable software products (i.e. every feature combination allowed according to the feature model) to detect feature interaction, \textit{pairwise feature-interaction testing} can be applied in the context of SPLs \citep{Oster.2011}.
In pairwise feature-interaction testing a subset of all derivable software products is generated in that every feature pair is comprised.
For a better feature interaction detection, works such as \citep{Cohen.2008} discuss a more general approach of t-wise testing, i.e. testing software products that contain every combination of \textit{t} features, with \textit{t} $\in$ \{1,2,3,...,n\} representing the coverage strength .
However, most researchers propose pairwise feature-interaction testing approaches \citep{Sahid.2016}. 

In SPLs, the resolution of feature interaction cannot be hard-coded as during product derivation the features are selected individually obeying the constraints set out by the feature diagram \citep{Hunt.2007}.
However, additional code that resolves the unintended behaviour may be conditionally included when two interacting features are present in a software product as proposed for the \textit{optional-feature problem}.
If a feature contains code that depends on another optional feature, the second feature becomes mandatory contrary to its specification, which is known as the optional-feature problem \citep{Kastner.2009}.
Different approaches propose extracting the dependent code into a separate module, which is called derivative \citep{Liu.2005,Liu.2006} or lifter \citep{Prehofer.1997}.
Consequently, both features may be included independently and used in software products.
However, when using both features in combination, the \textit{derivative module} is included as well.
A derivative module constitutes the resolution of technical dependencies and is not a feature.
Consequently, it is not added in the feature model \citep{Kastner.2009}.
Furthermore, in order to resolve the interaction of two features they may be marked as mutually exclusive in the feature model, or precedences or priorities may be defined.
For a review of resolution techniques for feature interactions we refer interested readers to \citep{Soares.2018}.

\subsection{FeatureIDE}
\textit{FeatureIDE} \citep{Thum.2014} is an Eclipse-based integrated development environment (IDE) for SPLs. It covers the four phases of Software Product Line Engineering set out in Section \ref{subsect:phases}.
To support domain analysis, FeatureIDE provides a graphical editor to model features and their dependencies in a feature model.
Feature models contain constraints, are stored as XML files, and can be imported to or exported from FeatureIDE.
Furthermore, FeatureIDE provides the user with convenient refactoring tools and inconsistency detection.

FeatureIDE assists the developer with a configuration editor to create a configuration for a specific software product during requirement analysis.
The developer may mark those features from the feature diagram to be contained in the software product.
The configuration editor prevents configurations not obeying the constraints of the feature diagram.
Finally, the configuration is stored in a configuration file.

FeatureIDE supports various frameworks for implementing features during domain implementation.
This includes FeatureHouse and the AHEAD tools.

For software generation, FeatureIDE takes a configuration file as input and composes the software artefacts corresponding to the selected features.
FeatureIDE places the composed software artefacts in a specified output directory.
The artefacts in the output directory are then natively compiled as artefacts of the language the framework for the domain implementation is based on.

\section{Previous Work}
\label{sec:background}
In our previous work, we introduced the approach of \textit{Process-Aware Information System Product Line} (PAIS Product Line) \citep{Hehnle.2023}.
This approach applies the concepts of SPL Engineering to the development and maintenance of similar PAISs in order to reduce development efforts and costs by enabling the reuse of common core artefacts.

\subsection{Variation Points}
A PAIS Product Line constitutes a common set of core artefacts including a core process model from which PAIS products may be derived.
Various activities of the core process model may be declared as variation points, i.e. different implementations may be selected for these activities during PAIS Product derivation.
Throughout this work, BPMN 2.0\footnote{\url{https://www.omg.org/spec/BPMN/2.0.2/PDF}} is used as notation to model business processes as we agree with other researchers that BPMN is the de-facto industry standard for process modelling \citep{Dohring.2011,HongyanZhang.2014,Delgado.2022}.
In addition, BPMN process models can be executed by a process engine, which is harnessed by our proof-of-concept.
In BPMN 2.0, an activity corresponds to a step on which work is performed, which is either atomic or compound.
While call activities (call subprocess, which itself comprise activities) constitute compound activities, tasks are atomic activities that cannot be further broken down.
For the sake of this approach, the BPMN 2.0 \textit{send task, script task, service task,} and \textit{business rule task} are regarded as \textit{automated tasks} as they all represent an automatic step in a business process without user interaction.

Neither \textit{manual tasks} nor the \textit{receive tasks} are considered by this approach as \textit{manual tasks} do not occur in digitized business processes and \textit{receive tasks} may be substituted by \textit{receive events}. 
A BPMN 2.0 \textit{user task} corresponds to a step in a business process where the user interacts with the PAIS. Consequently, the approach considers the following three activity types: \textit{call activity, automated task,} and \textit{user task}.

\begin{figure}
	\centering
	\vfill
	\includegraphics[width=80mm]{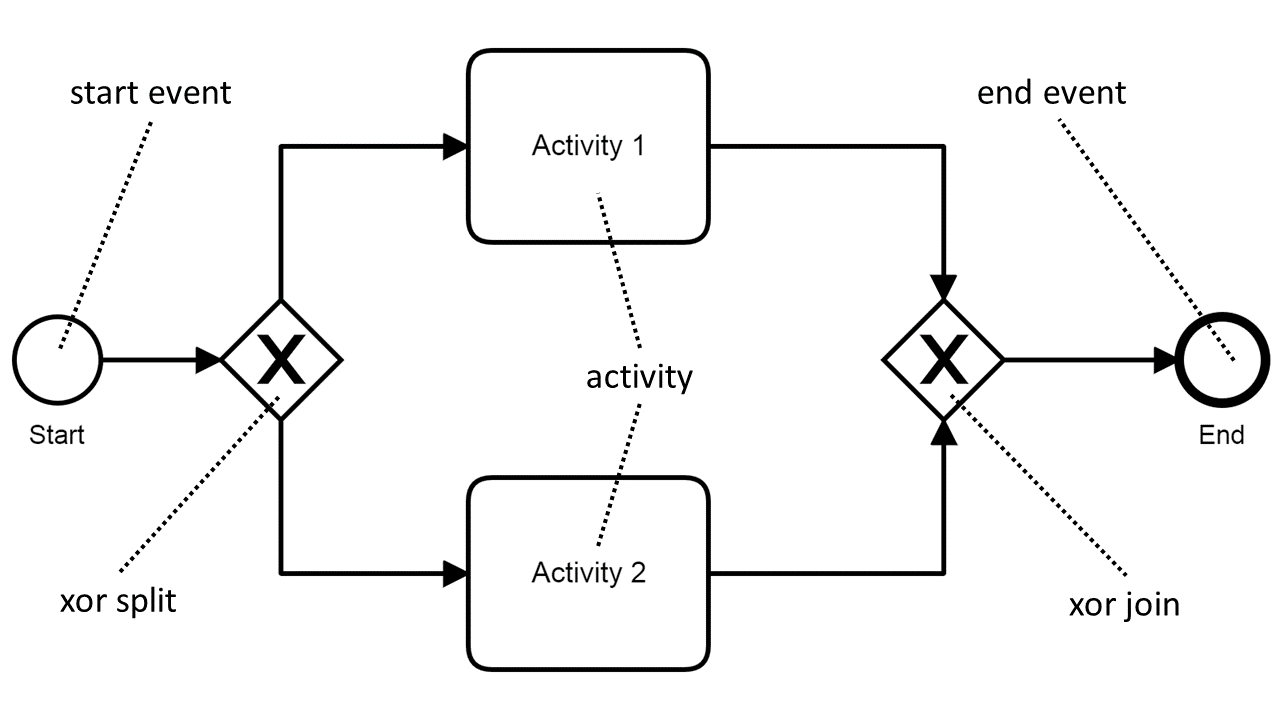}
	\caption{Example Business Process \citep{Hehnle.2023}} \label{example-process}
	\vfill
\end{figure}

Figure \ref{example-process} shows an example business process comprising two activities, whose activity types are not specified, i.e.
the activities constitute the variations points.
During PAIS product derivation, an implementation will be selected for these activities and hence the activity types be determined.
Depending on the selected implementation, the activities might be automated activities (i.e. executing business logic), user tasks (i.e. a user form) or call activities (i.e. calling a subprocess).
An activity might also remain without a type, i.e. no implementation is selected.
Consequently, the activity is neither a call activity nor an automated task nor a user task.
An untyped activity has no function in a PAIS.
A process instance that passes an untyped activity does nothing, i.e. neither a user form is invoked nor automated logic is executed.

\subsection{Variability Modelling}
\begin{figure}
	\centering
	\vfill
	\includegraphics[width=100mm]{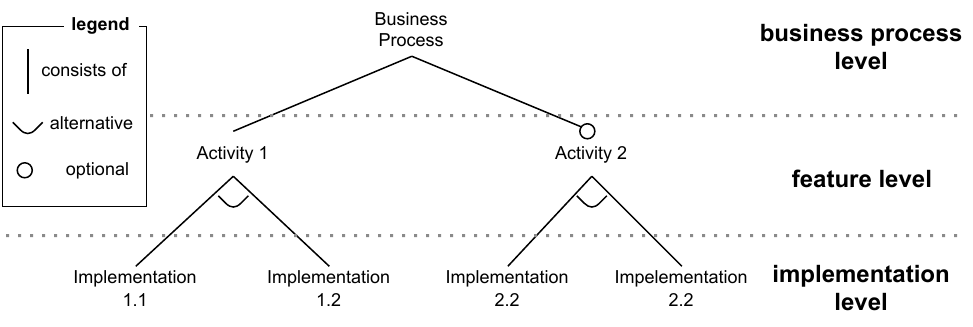}
	\caption{Process Feature Model \citep{Hehnle.2023}} \label{fm-pais}
	\vfill
\end{figure}

The approach transfers the concept of features from SPL Engineering to process activities.
This enables us to create \textit{process features models}, which depict what implementations may be selected for which activity, and which activities are optional and which mandatory.
The process feature model depicted in Figure \ref{fm-pais} reflects the process model from Figure \ref{example-process}.
In the process feature model, for each of the two activities one out of two implementations may be selected.
Furthermore, Activity 2 is optional, i.e. an implementation may be selected, but this is not mandatory.

A process feature model consists of three layers, i.e.. the process level, the feature level, and the implementation level.
The process level represents the entire business process.
The feature level includes all activities of the business process and specifies whether they are mandatory (i.e. an implementation has to be selected) or optional (i.e. an implementation may be selected but this is not mandatory).
The implementation level lists the available implementations of the activities.
If an activity corresponds to a subprocess, another process feature model with the same structure (3 levels) has to be constructed for this subprocess.

\subsection{Feature Interaction}
Each activity of a business process has input and output data whose data structure is specified at design time.
The term \textit{process data structure} is used to refer to the combined input and output data of all activities of a business process.
Thus, the input/output data structure of an activity determines that part of the process data structure to which the activity has read/write access.
Two activities with the same output data structure have the same write access.
Hence, an activity that has the same output data structure as its preceding activity might overwrite the data of its predecessor, i.e. the data of the predecessor is lost.
During requirements engineering, the order of the activities as well as their output data structure (i.e. write access) is specified.
Therefore, the behaviour of the activities can be foreseen and data will not inadvertently be overwritten.
Aside from passing on and accepting data, there is no \textit{unintended} interaction between the activities as they are executed in a predefined order.
If two activities in a process are modelled in parallel (e.g. after a gateway), they are often still executed sequentially by common Workflow Management Systems (WfMS) by default.
When configured accordingly, parallel modelled activities may be executed in different threads to achieve concurrency.
While there are some ways in which the threads may interact from a technical perspective, e.g. accessing shared memory, pausing one thread in order to wait for the completion of another, these interactions would have to be developed explicitly and would therefore be intended. 
But, in general, these kinds of interactions do not reflect the requirements of a business process.
Reading shared memory or other interactions between activities modelled in parallel (i.e. executed in parallel threads) predominantly contradict the requirements of business processes.
Activities of a business process correspond to logically isolated features, which in general read process data, transform the data (e.g. by calculations or by prompting for user input) or call web services, and write the results back to the process data.
Write access to process data needs to be considered during design time as mentioned previously.
Consequently, there are no \textit{unintended} feature interactions in PAIS Product Lines. 		
\section{Flexibility for PAIS Product Lines}
\label{sec:approach}
The approach of PAIS Product Lines currently lacks flexibility. Only one implementation can be chosen for an activity during PAIS Product derivation. 
Furthermore, the process data structure cannot be adapted to the individual need of an organisation. All derived PAIS Products share the same process data structure.
Finally, the approach of PAIS Product Lines currently supports feature binding at compile time, solely.

This section shows how to eliminate the limitations to flexibility by enhancing the approach of PAIS Product Lines.
First, the requirements of the enhanced approach are elicited.
Then, the assumptions for the enhanced approach are stated.
Finally, it is shown how the aforementioned limitations concerning flexibility can be eliminated.
 
\subsection{Requirements}
\label{sect:requirements}

The requirements for our approach were deduced from a cooperation with German municipalities.
German municipalities have organised various events at regional and state level (See Table \ref{tab:workshops} for events we attended) to share experiences in the digitisation of municipal business processes involving various stakeholders, including the respective state government, digitisation companies from the public and private sectors, and municipal umbrella organisations.
During these events, experiences are exchanged on approaches and technical, legal, and financial challenges in the digitisation of municipal business processes.
Due to our knowledge in business process management and digitisation, one co-author and two co-workers\footnote{The referenced co-author simultaneously held and holds part-time positions as a research associate and as an employee at a digitisation company. The mentioned co-workers were employed by the same firm.} were invited to attend three events (cf. Table \ref{tab:workshops}) and act as facilitators during the breakout sessions.
In the following, we describe the three events and our role.

In Event 1, there was a one breakout session workshop (90 minutes), in which a task force of three municipalities discussed the requirements of the business process \textit{special parking permit for craftspersons}.
The objective of the task force was to agree on common requirements for three business process so that uniform PAISs could be developed while reducing redundancies and costs.
Each municipality was supposed to digitise one business process and share it with the partner municipalities, so that each municipality only had to digitise one business process but could use three.
We acted as facilitators in the breakout session in which the requirements for the business process \textit{special parking permit for craftspersons} should be aligned.

A variety of collaboration and participation approaches have been proposed to elicit requirements and to model business processes in joint sessions by involving different stakeholders (e.g. end users) to include diverse perspectives \citep{Barjis.1211201112142011,Lai.05262014,Front.2017}.
While these approaches focus on different methods with varying level of detail (e.g. incorporating group storytelling or role plays), all have joint sessions in common in which domain experts present the requirements verbally, a facilitator mediates between them, and a modelling expert designs the business processes.
Although we did not apply any specific approach, we followed the basic idea of these approaches.
As opposed to the mentioned approaches, we were facing the challenge to not only involve different stakeholders of the same business process but rather stakeholders of multiple similar yet slightly different business processes.

In line with the roles described by \cite{Lai.05262014}, at least one domain expert (i.e. municipal clerk who is a process participant) and an IT expert from each municipality were present to share the expectations and activities of the business process.
We acted as facilitators and process modellers to collect the requirements and model the business process graphically.
Our main responsibility was to mediate between the different requirements of the municipalities, reach an agreement and thereby model a standardised business process.
While available software tools \citep{Barjis.1211201112142011,Front.2017} or even interactive tabletops \citep{Doweling.2013} support collaborative and participative process modelling, we relied on whiteboards and sticky notes like \cite{Lai.05262014}.
While we acknowledge that under normal circumstances 90 minutes are not sufficient to elicit the requirements for a business process, in this case, the three municipalities already had elicited the requirements of their respective business process.
Consequently, during the session, the requirements of the three municipalities only had to be collected, compared, and synthesised, and areas of alignment and divergence had to be identified.
After the event, we tried to make compromise proposals by phone and email.
However, despite our efforts, all compromise proposals were rejected without further workshops.
Consequently, the differences in the requirements proved impossible to develop a common PAIS.

Significantly more municipalities attended Event 2 to align their digitisation strategies and form collaborations at the state level.
While the event had a fixed agenda with presentations (e.g. by a representative of the interior ministry of the corresponding state government), the event was designed to provide room for networking and informal exchanges between municipalities, government institutions, and private companies supporting the digitisation endeavours of the municipalities and state government.
During informal conversations with further municipalities, we could confirm that the requirements regarding implementation variability of business processes we deduced for the special parking permit also apply to other municipal business processes.

Finally, in breakout sessions during Event 3, once again we acted as mediators and process designers to collect and compare the requirements of other municipal business processes, which include \textit{enrolling children in kindergarten} and \textit{applying and issuing the civil status documents for a citizen}.

In summary, we could specifically elicit the requirements for the business process \textit{special parking permit for craftspersons} of three German municipalities and deduce the challenge of implementation variability of business processes.
Furthermore, we could confirm that this challenge is not restricted to this business process but can be applied to other municipal business processes.

\begin{threeparttable}
	\centering
	\setlength{\leftmargini}{0.4cm}
	\begin{tabular}{ m{0.5cm}  m{0.5cm}  m{6cm}  m{7.5cm} }
		\textbf{ID} & \textbf{D\tnote{1}} & \textbf{Important Agenda Topics}  & \textbf{Participants}   \\ \hline
		1              & 8             & 
		\begin{itemize}[noitemsep,topsep=0pt]
			\item Exchange of digitisation approaches
			\item Legal challenges
			\item Technical challenges
			\item Discussions about joint digitisation of concrete business processes in break-out sessions in smaller groups\tnote{2}
		\end{itemize}
		
		&  
		\begin{itemize}[noitemsep,topsep=0pt]
			\item 17 municipal employees (8 municipalities) 
			\item 2 publicly owned corporations for digitisation
			\item 2 representatives of the interior ministry of the state government
			\item 1 co-author and 1 co-worker as mediator
			\item 3 representatives of another digitisation company
		\end{itemize}
		
		                      \\ \hline 
		2              & 6             &
		\begin{itemize}[noitemsep,topsep=0pt]
			\item Legal and financial conditions
			\item Discussion about future collaboration between municipalities, companies, and state government
			\item 3 breaks for informal exchanges
		\end{itemize}
		
		 &  
		\begin{itemize}[noitemsep,topsep=0pt]
		 	\item around 150 workshop participants including representatives from municipalities, companies (including co-author and co-worker as guests), publicly owned corporations for digitisation, and state government 
		\end{itemize}
		 
		 \\ \hline
		3              & 6             &
		\begin{itemize}[noitemsep,topsep=0pt]
			\item Benefits of digitisation
			\item Prioritisation for the digitisation of municipal business processes
			\item Break-out sessions to discuss commonalties and differences of concrete municipal business processes\tnote{2}
		\end{itemize} &   
		\begin{itemize}[noitemsep,topsep=0pt]
			\item 14 municipal employees (10 municipalities)
			\item 1 employee of a municipal umbrella organisation
			\item 1 employee of a publicly owned corporation for digitisation
			\item 1 co-author and 2 co-workers as mediators and process designers
		\end{itemize}
		                     \\ \hline
	\end{tabular}
	\begin{tablenotes}
	\item[1] Duration in hours excluding preparing and cleaning up the meeting rooms but including the coffee and lunch break, in which the topics were informally discussed.
	\item[2] Collaborative requirements elicitation facilitated by co-author/co-workers
	\end{tablenotes}
	\caption{Events for Digitising Municipal Business Processes}
	\label{tab:workshops}
\end{threeparttable}

Example \ref{xmpl:bp-craftsperson} describes the business process \textit{special parking permit for craftspersons} according to the requirements elicited during the conducted workshop, which, according to our findings, represent general widespread challenges in the digitisation of business processes in German municipalities.
In \citep{Hehnle.2023}, we used the business process in a simplified version as an example.

\begin{figure}
	\centering
	\vfill
	\includegraphics[width=120mm]{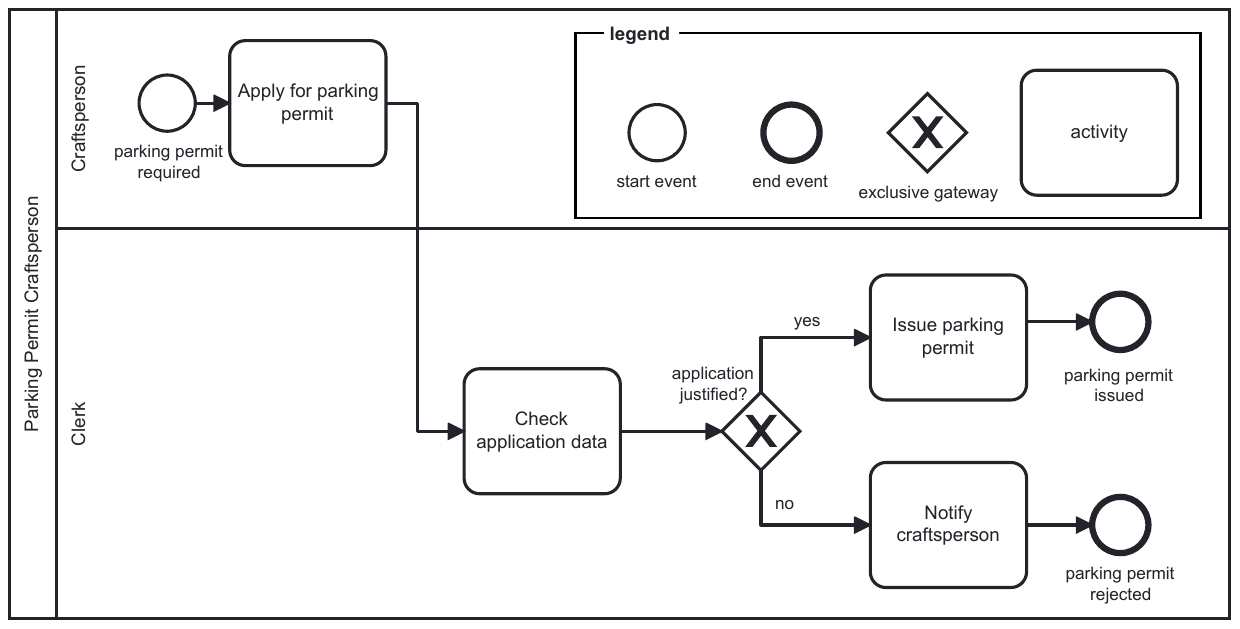}
	\caption{Business Process \textit{Special Parking Permit for Craftspersons}} \label{example-process-approach}
	\vfill
\end{figure}

\begin{customExample} %
	\label{xmpl:bp-craftsperson}
	(Special Parking Permit Process): Various German municipalities offer a special parking permit for craftsperson who may park their cars in zones where regular citizens need to pay or where parking is prohibited.
	The business process of checking an application for this parking permit is the same among all considered municipalities and is displayed in Figure \ref{example-process-approach}.
	First, a craftsperson applies for the parking permit.
	Then, the application is checked.
	If the application is justified, the parking permit is issued.
	If the application is not justified, the applicant is notified of the rejection.
	
	While the described control flow is common across all municipalities, the implementation of the business process varies.
	In some municipalities, the application check is carried out by a municipal clerk whereas in other municipalities the application is checked automatically by comparing the data of the application with the data of the craftsperson who is officially registered with the authorities.
	Besides, when the application is rejected the applicant may be notified via e-mail, or SMS, or by a municipal clerk depending on the municipality and the choice of the applicant.
	The applicants might choose the way they will be notified from a list of available notification means.
	
	Furthermore, the process data structure also varies as the municipalities require the applicant to fill in different data.
	In some municipalities, the parking permit is valid for every car whereas in others the applicant needs to provide the number plate for which the parking permit shall be valid in order to prevent misuse of the parking permit for the craftsperson's private car.
	In addition, to be able to automatically check the applications, the craftspersons need to provide the commercial register numbers under which they are registered with the authorities.
\end{customExample}

In a nutshell, PAIS Product Lines need to comprise common core artefacts including a process model from which similar PAIS Products can be derived in order to reduce the development efforts in comparison to developing the similar PAIS Products separately.
One or more implementations shall be selectable for an activity that constitutes a variation point.
The implementations shall be selectable at compile time, start time, and runtime and the process data structure shall be customisable for each PAIS Product as well.
The enhanced approach of PAIS Product Lines shall meet the following requirements:

\begin{enumerate}[label=R\arabic*:,  font=\itshape]
	\item In line with our previous work, PAIS Products need to be derivable from a PAIS Product Line which denotes a set of common core artefacts including a core process model in order to avoid the development of redundant software artefacts for similar PAISs.
	\item Multiple implementations for one activity shall be selectable. Only the selected implementations shall be contained in the derived PAIS to keep the size of the latter to a minimum.
	\item During PAIS Product derivation the process data structure shall be customisable allowing different organisation to use different process data structure in their derived PAIS Product.
	\item It needs to be formally specified what implementations for which activities may be selectable and what constraints need to be met during this selection.
	\item Combining certain implementations of different activities might produce unintended behaviour (feature interaction). Methods need to be established in order to detect and prevent unintended feature interactions.
	\item To enable selecting activity implementations during start and runtime based on input of users' dynamic feature binding becomes necessary.
	When using dynamic feature binding, the challenges described in Section \ref{subsect:binding} need to be tackled:
	\begin{enumerate}
		\item When an activity implementation is dynamically unbound during runtime (i.e. the implementation is deselected) the data it was processing might get lost.
		It needs to be specified whether an activity implementation may be unbound and - if yes - how, e.g. saving the current state.
		\item When unbinding an activity implementation during runtime, shared data and system processes might be affected.
		Methods need to be established ensuring that shared data and system processes will not be corrupted.
		\item When using static and dynamic feature binding, activity implementations might be bound in a different order than intended.
		It must therefore be assured that different binding orders must not lead to unintended behaviours.
	\end{enumerate}
	
\end{enumerate}

\subsection{Assumptions}

Compared to previous work \citep{Hehnle.2023}, we have lifted all but two assumptions:

\begin{itemize}
	\item The control flow of the core process model included in a PAIS Product Line is specified and immutable, i.e. the process model does not change during PAIS Product derivation. Hence, all derived PAIS Products will have the same process model and thus an identical control flow. While this work focuses on implementations variability, other approaches deal with control flow variability \citep{vanderAalst.2006,AlenaHallerbach.2008,Hallerbach.2010,Hallerbach.2010b}.
	The requirements we elicited with the three German municipalities for the business process \textit{special parking permit} do not require control flow variability.
	Nevertheless, based on informal discussions with further German municipalities we are aware of cases in which both implementation variability and control flow variability of business processes are necessary.
	However, we expect to solve lots of cases using implementation variability, only. 
	Future work shall reveal how to combine implementation variability and control flow variability to cover all cases of business process variability.
	\item The concepts will be applied to activity-centric business processes. 
	There are different approaches to business process management with each having advantages and disadvantages.
	While object- and data-centric approaches like those discussed by \cite{Kunzle.2009} and \cite{Steinau.2019} have some advantages, we will apply our concepts to BPMN as the de-facto industry standard (i.e. activity-centric business processes) to achieve broad acceptance.
\end{itemize}

\subsection{Multiple Implementations of an Activity}
\begin{figure}
	\centering
	\vfill
	\includegraphics[width=100mm]{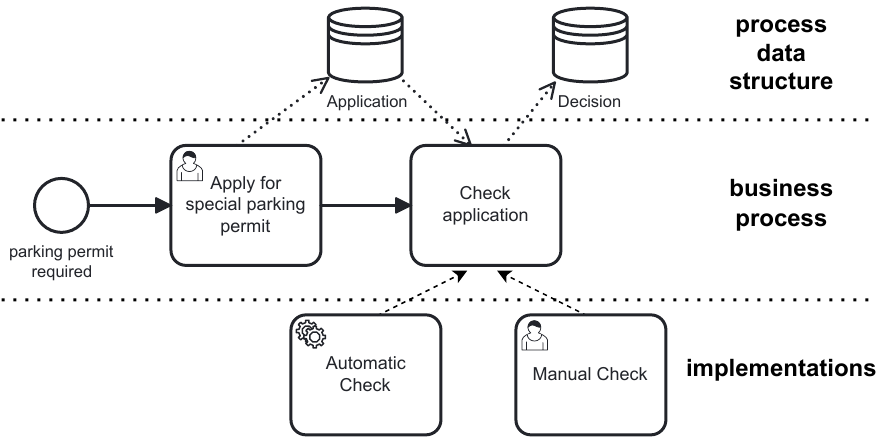}
	\caption{Extract of Business Process \textit{Special Parking Permit for Craftspersons} with Implementation Options} \label{example-process-part}
	\vfill
\end{figure}

In our previous work, when deriving a PAIS Product, for each activity constituting a variation point only one implementation may be selected, i.e. the implementation may be selected if this activity is either a manual task consisting of a user form or an automated task executing business logic.
In Figure \ref{example-process-part} an extract of the business process special parking permit is depicted.
Activity \textit{Check application} has no type and hence corresponds to a variation point.
Two implementations are available, i.e. \textit{Automatic Check} and \textit{Manual Check}.
In our previous work, only one of the two implementations may be selected.
Furthermore, the activities of a business process are executed in a pre-specified order, e.g. activity \textit{Check application} is executed after activity \textit{Apply for special parking permit}.
The execution order as well as the read and write access of each activity to attributes of the process data structure is determined during requirements engineering.
In Figure \ref{example-process-part}, it can be seen that activity \textit{Apply for special parking permit} has write access to the \textit{Application} data structure, whereas activity \textit{Check application} has read access to \textit{Application} and write access to the \textit{Decision} data structure.
Consequently, the activities only influence each other with the data they read and write. 
As the read and write access is carefully specified there are no unintended feature interactions.

Building on previous work, PAIS Products are derived from a PAIS Product Line consisting of a set of common core artefacts including the process model and the activity implementations (cf. Requirement \textit{R1}).
When allowing for multiple implementations of an activity, no execution order of these implementations is specified.
If, for example, in Figure \ref{example-process-part}, for activity \textit{Check application} both implementations \textit{Automatic Check} and \textit{Manual Check} are selected, both implementations run in parallel when the process instance passes this activity.
As the implementations of an activity may be executed in isolation there is no unintended technical feature interaction.
However, implementations of one activity may influence each other via the data they write.
Due to non-determinism, it might not be foreseeable which implementation finishes last and thereby overwrites the data of its predecessors, i.e. the results are unpredictable.
Consider Example \ref{xmpl:read-write}.

\begin{customExample}
	\label{xmpl:read-write}
	(Feature Interactions in PAIS Product Lines): In Figure \ref{example-process-part}, activity \textit{Check application} writes the decision on whether an application is justified to the process data structure.
	Consequently, both implementations may write the decision as well, i.e. the application is checked both automatically and by a municipal clerk.
	When selecting both implementations (i.e. manual and automatic) for activity \textit{Check application}, the implementation that finishes last determines whether the application is justified.
	For example, the result of the automatic application check is overwritten by the decision of the municipal clerk if the clerk takes longer than the automatic application check.
	This leads to unpredictable results depending on which implementation writes the data last.
	This can be seen as unintended feature interaction.
\end{customExample}

To tackle this challenge, activities reading and those writing need to be distinguished.
Activities that have read-only access to data do not pose a challenge in terms of feature interaction.
In Example \ref{xmpl:bp-craftsperson}, when selecting both SMS and e-mail for notifying the applicant it does not matter what implementation finishes first as the implementations do not write data and hence do not influence each other.
In order to prevent feature interactions in activities that write data, prevention techniques from SPL Engineering might be taken into consideration.

In Section \ref{sect:feature-interaction}, mutual exclusion of the interacting features, precedences and priorities are discussed as solutions to prevent feature interactions.
However, due to Requirement \textit{R2} it becomes necessary to be able to co-select multiple implementations for one activity, mutual exclusion cannot be used to prevent feature interaction.
Specifying precedences and priorities means that the implementation with the highest priority will always overwrite the data of the other implementations.
Consequently, there is no sense in including multiple implementations when the one with the highest priority always overwrites the data of the others.

Another option would be to not grant the different implementations write access to the same data.
This does not work either as the implementations implement the same business requirement in different ways and therefore need to write the same data.
For example, the application check implementations all need to write the data on whether or not the application is justified.
Note that the concept of derivative modules (cf. Section \ref{sect:feature-interaction}) may be used in this context as well.

Additional code can be included when two implementations write to the same data structure.
For example, additional code could determine that the application is not justified as soon as at least one of the two implementations decides that the application is not justified.
This additional code aggregates the data from the different implementations.
We refer to this additional code as \textit{aggregation code}.
Another aggregation code might constitute a majority vote.
If there are multiple application check implementations, the application is rejected when a majority of the implementations consider the application being unjustified.
As opposed to derivative modules, aggregation code needs to be placed in the process feature model as aggregation code does not solely resolve technical dependencies.
Aggregation code introduces resolution on a business level.
There might be multiple aggregation codes from which one may be selected during PAIS product derivation.
This choice needs to be indicated in the process feature model.
In the process feature model, all implementations of one activity are grouped as well as all available aggregation codes for that activity.
A dashed arrow introduces the constraint that one aggregation code needs to be selected when more than one implementation of an activity is selected.
Consider the following example.

\begin{customExample}
	\label{xmpl:feature-model}
	(Aggregation Code): Figure \ref{feature-model-approach} shows the process feature model for Example \ref{xmpl:bp-craftsperson}.
	For activity \textit{Check application} two implementations may be selected, a form for a manual check and/or an automatic check.
	In addition, two aggregation code implementations are available.
	The feature model notation described in Section \ref{sect:var-model} was extended in Figure \ref{feature-model-approach}.
	The two implementations for activity \textit{Check application} as well as the two aggregation code implementations are each grouped using a dashed box.
\end{customExample}

A dashed arrow connects the implementation group \textit{I} and the aggregation code group \textit{A} of the same activity.
Its label states 
\[ \#I > 1 ? requires \#A=1 \]
which reads as follows: If the number of selected implementations is greater than one, one aggregation code implementation needs to be selected, i.e. if two items from Group I are selected, one item from Group A needs to be co-selected as well.
The dashed arrow, which represents a conditional \textit{requires}, was taken from literature (cf. Section \ref{sect:var-model}) and supplemented with a label.
Using aggregation code, it becomes possible to prevent unintended interaction when selecting multiple implementations for one activity (cf. Requirements\textit{R2} and \textit{R5}).

\subsection{Varying Data Structures}
In our previous work, during PAIS Product derivation the process data structure could not be customised to the individual needs of an organisation.
All derived PAIS Products shared the same process data structure.
However, Example \ref{xmpl:bp-craftsperson} describes a scenario in which it becomes necessary to adapt the process data structure during PAIS Product derivation.

\begin{customExample}
	\label{xmpl:varying-data-structure}
	(Varying Process Data Structure in a PAIS Product Line): When applying for the special parking permit, some municipalities require the applicant to fill in the number plate whereas the parking permit of other municipalities is valid for all vehicles it is placed in.
	Furthermore, to automatically check the application, the application requires the commercial register number.
	Consequently, the process data structure varies between the municipalities.
\end{customExample}

As can be seen in Example \ref{xmpl:varying-data-structure}, different organisations need to customise the process data structure when deriving a PAIS Product from a PAIS Product Line.
Consequently, there are optional parts of the process data structure.
Activities read and write data.
The optional parts of the process data structure are accessed at least by some activity implementation or otherwise the selection of the former was useless as they are never used.
Hence, there are dependencies between activity implementations and the data structure.
Including the process data structure with the optional parts in the process feature model allows pointing out the dependencies between the activity implementations and the optional parts of the process data structure.
To point out the dependencies dashed arrows like the ones used in literature can be employed.
During PAIS Product derivation, when the activity implementations are selected, the corresponding optional parts of the process data structure need to be included in the PAIS Product in order to enable the compilation of the latter.
From a business requirements perspective, each included optional part of the data structure needs to be accessed at least once for writing and once for reading, in exactly this order.
Otherwise, if one part of the data structure is accessed for reading before writing, the data structure will be empty.
Furthermore, if one part of the data structure is accessed for writing, solely, there is no point in collecting the data at all as the data is never being processed or used.

\begin{figure}
	\centering
	\vfill
	\includegraphics[width=100mm]{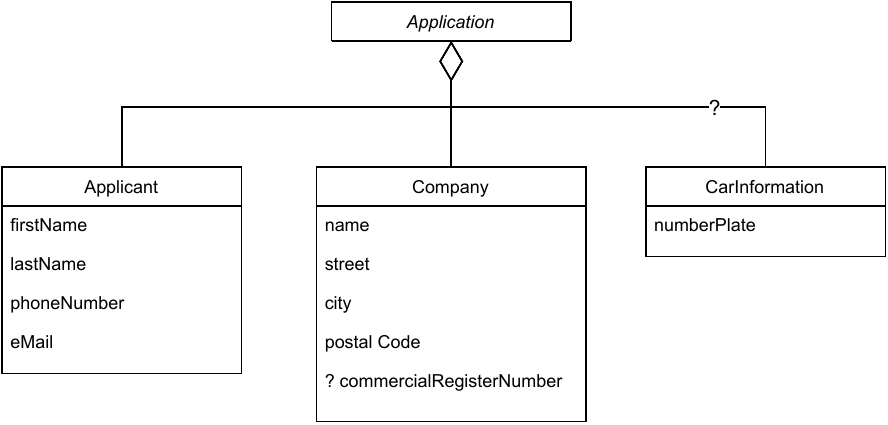}
	\caption{Data Structure of Example \ref{xmpl:varying-data-structure}} \label{data-structure-approach}
	\vfill
\end{figure}

In the following, the process of checking an application for the special parking permit (cf. Examples \ref{xmpl:bp-craftsperson} and \ref{xmpl:varying-data-structure}) is used to explain the concept of varying data structures.
Figure \ref{data-structure-approach} shows the process data structure associated with Example \ref{xmpl:varying-data-structure} as UML class diagram.
The \textit{application} corresponds to a composition of the data of the \textit{applicant}, the data of the \textit{company}, and the \textit{car information}, whereas the latter and the field \textit{commercial register number} in the data of the \textit{company} are optional (indicated with question marks).
The optional parts of the data structures are only included if necessary (cf. Requirement \textit{R3}).

\begin{figure}
	\centering
	\vfill
	\includegraphics[width=120mm]{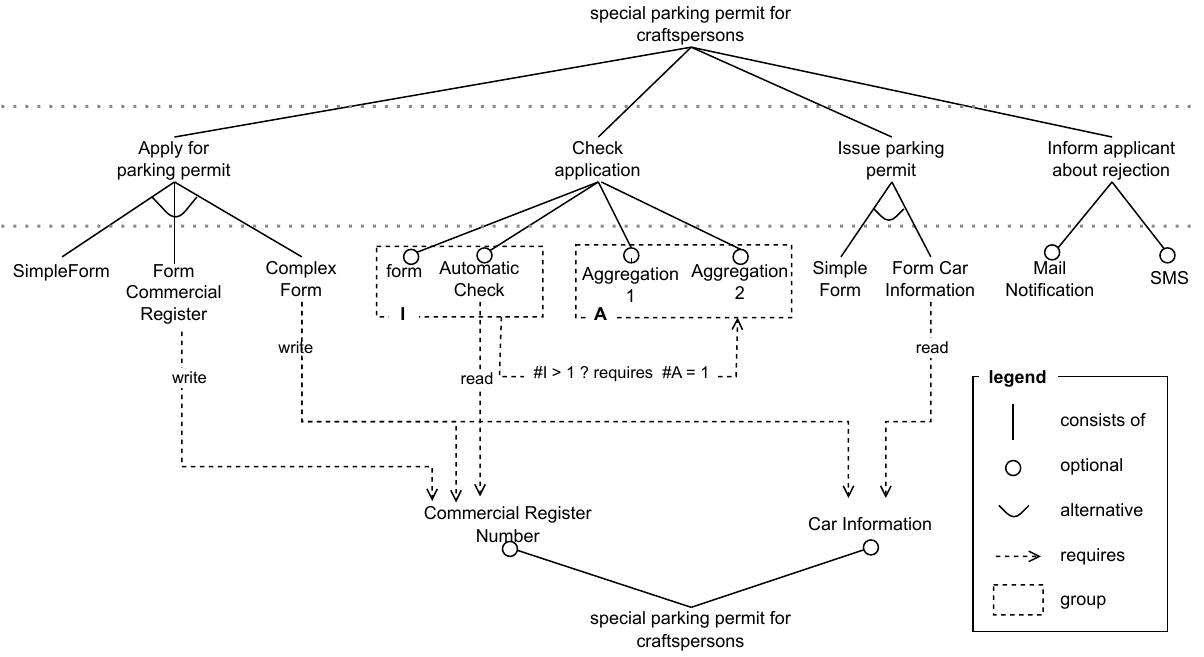}
	\caption{Process Feature Model} \label{feature-model-approach}
	\vfill
\end{figure}

Each activity implementation requires access to a specific part of the data structure, e.g. the automatic application check requires the commercial register number field in the data of the company.
In addition to the process feature model, bottom of Figure \ref{feature-model-approach} contains a feature model representing the data structure from Figure \ref{data-structure-approach}.
Those activity implementations that need a specific optional part of the data structure have dashed arrows to the required data structure part with a label indicating whether the implementation needs read or write access.
For activity \textit{Apply for parking permit} multiple alternative implementations exist.
There are three forms: a simple form, a form in which the applicant needs to fill in the \textit{commercial register number}, and a complex form that contains input fields for setting the \textit{commercial register number} and the \textit{number plate}.
All forms need write access to the corresponding data structure parts.
The automatic application check implementation needs read access to the \textit{commercial register number} to be able to compare the application data with the data from the commercial register.
For activity \textit{Issue parking permit}, there are two alternative forms.
One does not require optional parts of the data structure, whereas
the other form requires read access to the \textit{number plate} field as the number plate needs to be present on the issued parking permit. 

Using the feature model presented in Figure \ref{feature-model-approach}, the constraints between the implementations are the optional data structure fields become apparent (cf. Requirement \textit{R4}).

\subsection{Static and Dynamic Feature Binding}
The activity implementations are independent from each other and have no technical inter-dependencies.
The single point of interaction among the activity implementations is the write and read access to the process data structure.
Consequently, the order in which the activity implementations are bound does not influence their behaviour (cf. Requirement \textit{R6c}).

If an activity implementation runs during process execution, as it was bound during process start and selected by a user, the implementation will not be unbound during execution. 
Once selected for a process instance, an activity implementation will always finish it execution.
Hence, no data will be lost (cf. Requirement \textit{R6a}).

Shared data and system processes do not pose a problem as activity implementations are independent and have no technical dependencies.
Furthermore, activity implementations may not be unbound during process execution once they are selected (cf. Requirement \textit{R6b}).

\section{Proof-of-Concept}
To validate the presented approach, Example \ref{xmpl:bp-craftsperson} is implemented as a proof-of-concept, which is available at GitHub\footnote{\url{https://github.com/hehnle/process-implementation-variability}}.
For the proof-of-concept, an unaltered version of Camunda Platform\footnote{\url{https://docs.camunda.org/manual/7.22/}} is used as Workflow Management System (WfMS) embedded in a Java Spring Boot application.
In an effort to avoid proprietary software for our approach and guarantee seamless updates of Camunda Platform we did not customise Camunda Platform or other tools used for the proof-of-concept.
When customising tools directly or indirectly by developing plugins for a tool, these customisations and plugins need to be kept up to date and compatible with said tool.
As we are not able to maintain software sustainably, we solely use available tools as they are.

In the following, the properties are stated that need to be provided by the proof-of-concept.
Then, from a technical perspective the proof-of-concept shall show how to customize the process data structure during PAIS Product derivation, how to bind multiple implementations for one activity, and how to bind the implementations at different times.
Finally, it is portrayed how the development and maintenance may be supported by a tool.

\subsection{Properties}
In line with Requirements \textit{R1}-\textit{R6} (cf. Section \ref{sect:requirements}), the proof-of-concept needs to provide the following properties. In addition, the development phases of a PAIS Product Line shall be tool-supported similar to SPL Engineering, which results in \textit{P7}.

\begin{enumerate}[label=P\arabic*:,  font=\itshape]
	\item  PAIS Products are derivable from a PAIS Product Line. All derivable PAIS Products include the mandatory activity implementations, but may differ in the selection of alternative activity implementations. The proof-of-concept has to show how PAIS Products can be derived from a PAIS Product Line from a technical perspective, i.e. what variability mechanism is used.
	\item Multiple implementations of an activity might be selected and executed. A suitable mechanism needs to be chosen that allows to execute multiple implementations for one activity.
	\item A variability mechanism needs to be used to compose the process data structure during PAIS Product derivation to meet the organisations' individuals needs. 
	\item A process feature model outlines all activities (incl. implementations). Furthermore, it illustrates whether an activity is mandatory or optional.
	\item When using multiple implementations for one activity, their output data must be collected and aggregated in order to prevent the overwriting of data.
	This aggregation code needs to be conditionally included when more than one implementation of an activity is selected.
	\item In order to support both static and dynamic feature binding adequate variability mechanisms must be chosen.
	\item The phases domain analysis, domain implementation, requirements analysis, and product derivation shall be tool supported.
\end{enumerate}

\subsection{Data Structure Composition}
In our previous work, the build tool \textit{Apache Maven} was used to compose the artefacts during PAIS Product derivation.
The data structure was the same among all PAIS Products and the activity implementations were designed as independent logical units and were compiled JARs (Java applications) that were conditionally included when the implementation was selected based on \textit{Apache Maven} profiles.
That is, \textit{Apache Maven} was used to compose fully compiled artefacts.
Note that this worked, as these logical units were self-contained and could be put together like building blocks.
In contrast, when the process data structure needs to be customizable (cf. Example \ref{xmpl:varying-data-structure}), there are changes on the class level, i.e. before its compilation the process data structure needs to be customised.
Therefore, instead of composing JARs, Java classes need to be composed.
To compose Java classes, both FeatureHouse \citep{Apel.2009b} and the AHEAD tools \citep{Batory.2004} may be used.
As FeatureHouse does not require any additional key words like \textit{refines}, FeatureHouse is used to compose the Java classes of the process data structure in the proof-of-concept.

In Figure \ref{data-structure-approach}, class \textit{Company} has the optional field \textit{commercialRegisterNo} and the parent class \textit{Application} has an optional association with the class \textit{CarInformation}.

\noindent\begin{minipage}{.48\textwidth}
\begin{lstlisting}[language=Java, basicstyle=\tiny, frame=single, caption=Base Class ``Company'',
	label=list:base-class]
public class Company {

  private String name;
  private String street;
  private String city;
  private String zipCode;
}
\end{lstlisting}
\begin{lstlisting}[language=Java, basicstyle=\tiny, frame=single, caption=Classs ``Company'' Extension,
	label=list:extension]
public class Company {

  private String commercialRegisterNo;
}
\end{lstlisting}
\end{minipage}\hfill
\begin{minipage}{.48\textwidth}
\begin{lstlisting}[language=Java, basicstyle=\tiny, frame=single, caption=Composed Classs ``Company'',
	label=list:target]
public class Company {

  private String name;
  private String street;
  private String city;
  private String zipCode;
  private String commercialRegisterNo;
}
\end{lstlisting}
\end{minipage}

Listing \ref{list:base-class} shows the base Java class \textit{Company} with its attributes.
The extension of class \textit{Company} with the attribute \textit{commercialRegisterNo} is depicted in Listing \ref{list:extension}.
For the sake of brevity, the getter and setter methods of the classes in the listings are omitted.
The class composed by FeatureHouse from the base class and its extension is shown in Listing \ref{list:target}.
The base class, its extension, and the composed class are all pure Java classes.

\subsection{Multiple Implementations of an Activity}
Specific activities shall constitute a variation point, i.e. one or more implementations shall be selectable for these activities.
An implementation may be a service task, a user task, or a call activity, which calls a subprocess.
During PAIS Product derivation, the core process model needs to be composed with the implementations.
In this context, the implementations may be considered as refinements of the base process model.
The language Xak \citep{Anfurrutia.2007} can be used to compose the process models.
Note that the latter are modelled in terms of BPMN 2.0, which has an XML representation.

\begin{figure}
	\centering
	\vfill
	\includegraphics[width=100mm]{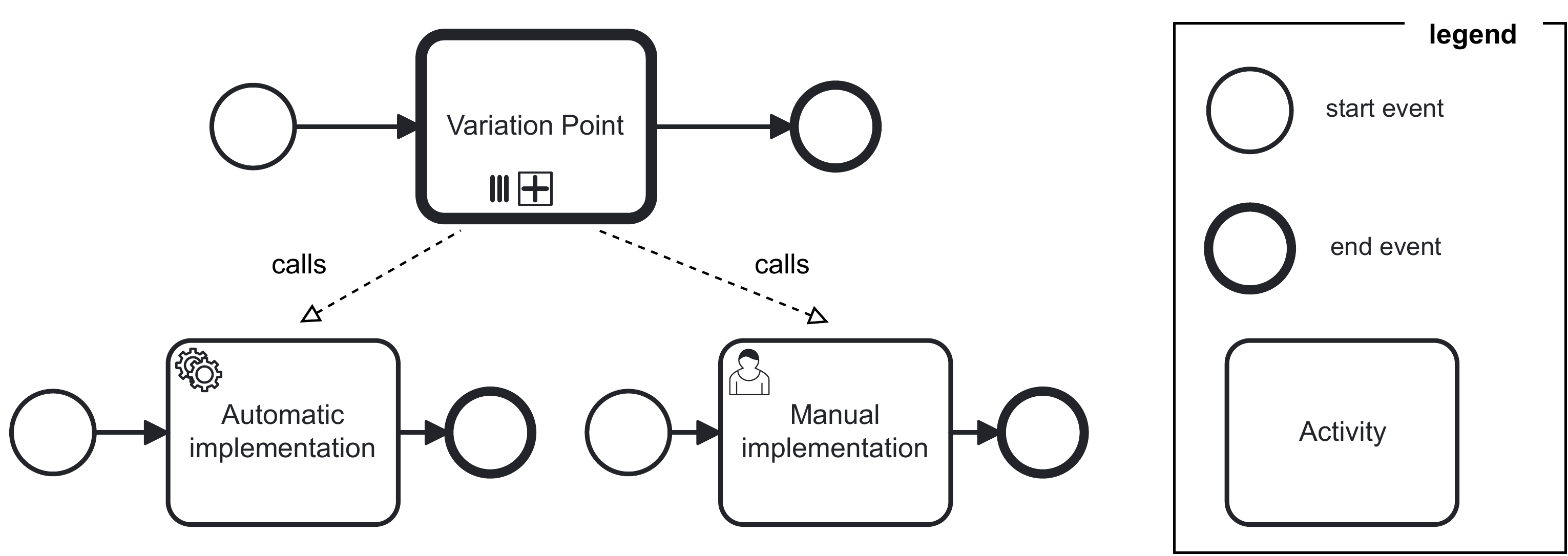}
	\caption{Call Activity Calling Multiple Implementations} \label{fig:call-activity}
	\vfill
\end{figure}

However, when using Xak, the refinements (i.e. the implementations) are not schema-compliant, i.e. the files in which an activity is refined (which are process models as well) cannot be edited in or displayed by BPMN 2.0 modelling tools.
As one advantage, BPMN 2.0 has both an XML and a graphical representation of the process model.
The graphical representation serves as a means of communicating process models between domain and IT experts.
If the process models serving as implementations are not schema-compliant, the advantage of BPMN 2.0, i.e. the graphical representation, does not come into effect.
Furthermore, when selecting lots of implementations these implementations need to be included in the core process model and the elements need to be rearranged in order to make room for the additional elements of the implementations.
Although works such as \cite{Scholz.2019} provide means for automatically layouting process models, with increasing number of included implementations the process model gets unclear and confusing.
Therefore, Xak is not used to compose the base process model and its implementations.
Instead, multi-instance call activities are used, which call the implementations as subprocesses.
The implementations (i.e. automated task or user task) are embedded in a process model, which is then called by the variation point activity.
Figure \ref{fig:call-activity} shows a simple process with a variation point activity which corresponds to a multi-instance call activity that calls two implementations (Note that the two dashed arrows are not part of BPMN but rather illustrate how a call activity calls subprocesses).
The implementations, in turn, are simple processes consisting of an activity, which is either an automated task or a user task.
As the variation point is a multi-instance activity it may invoke both implementations in parallel allowing for the execution of multiple implementations at the same time.

\begin{figure}
	\centering
	\vfill
	\includegraphics[width=120mm]{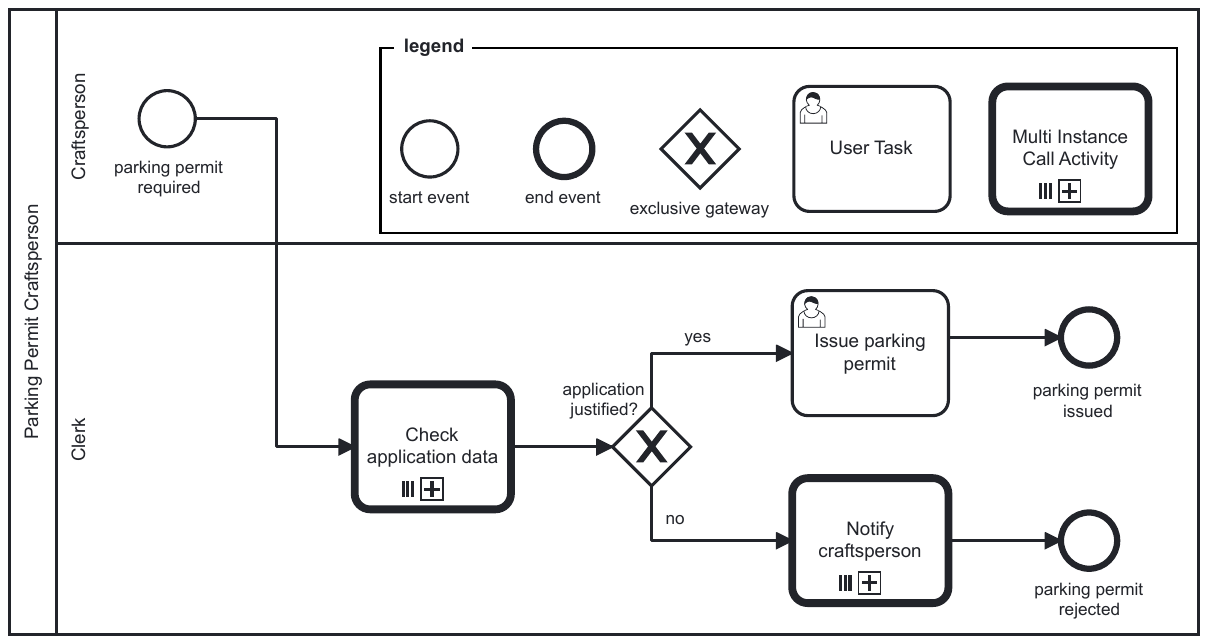}
	\caption{Business Process \textit{Special Parking Permit for Craftspersons}} \label{example-process-poc}
	\vfill
\end{figure}

Figure \ref{example-process-poc} shows the process model of Example \ref{xmpl:bp-craftsperson} for the proof-of-concept.
Activities \textit{Check application} and \textit{Notify craftsperson} are variation points, for which one or more implementations may be selected.
Therefore, these activities are multi-instance call activities, i.e. they call the corresponding implementations as subprocesses.
A detailed description on how the implementations are bound and called by the core process model is provided in Section \ref{sect:static-dynamic-binding}.
Note that activity \textit{Apply for parking permit} is removed in this process model.
Due to technical characteristics of Camunda Platform, the application form to apply for the parking permit is used to start the process and, hence, is not contained in the process model.

\subsection{Static and Dynamic Feature Binding}
\label{sect:static-dynamic-binding}
To be able to bind activity implementations statically as well as dynamically, approaches like the ones presented in Section \ref{subsect:binding} might be used.
Activity implementation code might be composed during compilation when the implementations are statically bound or language constructs (e.g. design patterns like decorators) are generated that allow choosing an implementation at runtime.
However, activities consist of graphical/XML representation and business logic (for service tasks) or HTML code (for user tasks).
Therefore, design patterns like the decorator pattern cannot be applied to the language mixture of a PAIS Product Line.
Instead, a plugin approach is used.
The activity implementations (e.g. automated task or user task) are embedded in a process model consisting of a start event, an end event and the actual activity implementation.
We refer to those process models embedding an activity implementation as \textit{implementation process models}.
Figure \ref{fig:call-activity} shows a core process model (at the top) with a variation point activity and two implementation process models (at the bottom) that can be invoked by the variation point activity.

\begin{figure}
	\centering
	\vfill
	\includegraphics[width=120mm]{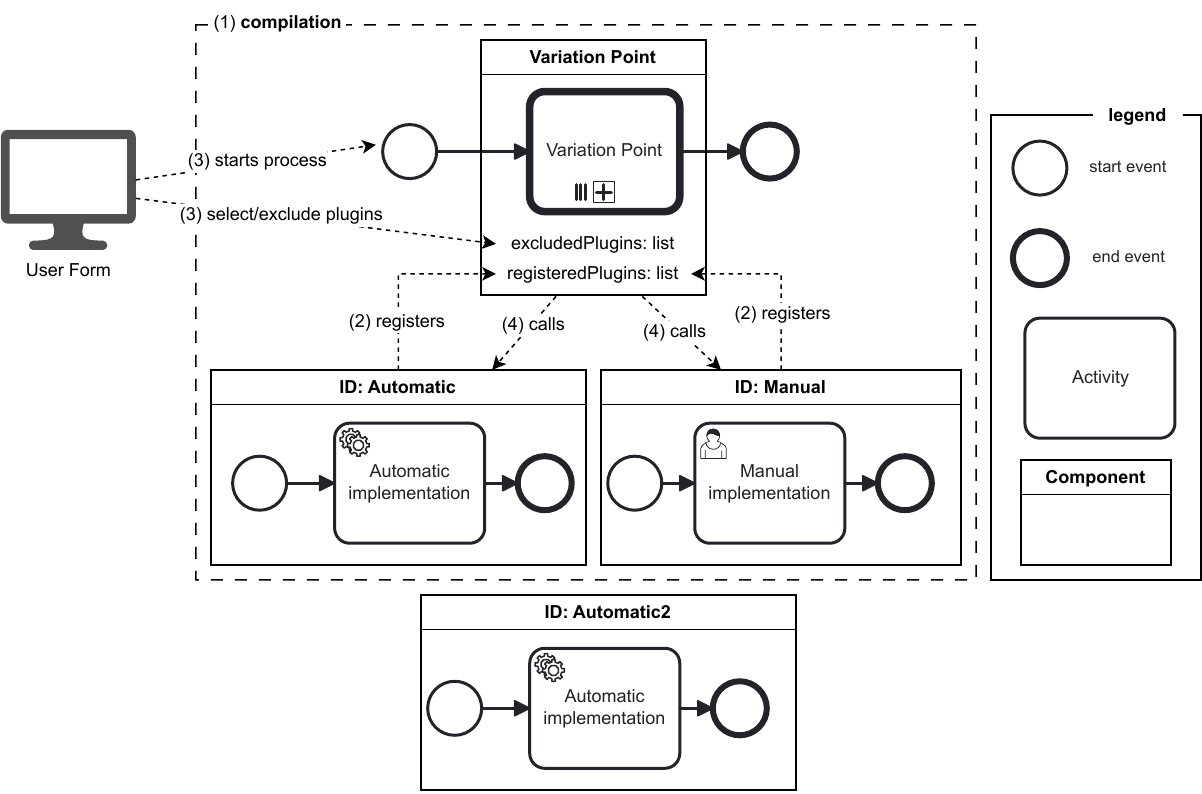}
	\caption{Process of Static and Dynamic Feature Binding in PAIS Product Lines} \label{fig:binding-pais}
	\vfill
\end{figure}

The plugin mechanism allowing for static and dynamic feature binding in PAIS Product Lines is illustrated in Figure \ref{fig:binding-pais}.
Figure \ref{fig:binding-pais} contains a process model with a variation point as well as multiple implementation process models that can be invoked by this variation point.
The processes are modelled using BPMN whereas solid boxes show in which components the process models are organised.
The dashed box and arrows describe the steps to achieve variability from compilation to runtime 
(Note that the boxes and arrows are not part of the BPMN specification).
During compilation, implementation process models may be included or excluded (static binding, cf. Figure \ref{fig:binding-pais} (1)).
Each implementation process model has a unique identifier (\textit{ID}), which is used to unambiguously identify the former.
When an implementation process model is included during compilation, it is registered as a plugin using its ID with the corresponding variation point activity (cf. Figure \ref{fig:binding-pais} (2)).
The excluded implementations cannot be registered as plugins and are not available at runtime.
Implementation \textit{Automatic2} in Figure \ref{fig:binding-pais} is not included during compilation and consequently not registered as plugin.
At start time, developers and, at runtime, users may deselect one or more implementations that have been registered for a variation point activity by setting parameters.
Figure \ref{fig:binding-pais} (3) shows a user form that can be used to start the process by selecting/excluding specific plugins (i.e. implementations).
Before a variation point activity calls the registered implementation process models it is checked whether one of the latter is excluded via a parameter (cf. Figure \ref{fig:binding-pais} (4)).
Only implementation process models are invoked that are not excluded via parameters (dynamic binding).

When selecting multiple implementations of a variation point activity that writes data, aggregation code needs to be included as well in order to prevent unintended overwriting of the same data.
As activity implementations are realised as implementation process models, aggregation code needs to map the data returned from the subprocesses (i.e. implementation process models) to the parent process model (i.e. the core process model) by aggregating and consolidating the data from the different implementations.

From a technical point of view, for static binding, FeatureHouse is used for composing the activity implementations (like for process data structure composition).
The core process is bundled with the implementation process models and the implementing logic, namely HTML code for user tasks and Java code for automated tasks as well as configuration files (i.e. application.properties files for Spring Boot).
As described in Section \ref{subsec:featurehouse}, FeatureHouse represents artefacts including their hierarchical structure as a tree and merges them.
That is, each feature corresponds to a folder containing the software artefacts within their hierarchical structure, i.e. the Java files in packages but also the process models and configuration files in their respective folders.
During composition, FeatureHouse merges the folders of the selected features taking into account the hierarchies within the folders (i.e. packages and folders).
In the proof-of-concept, each feature folder consists of the packages for the Java classes, a folder for the implementation process models, and when necessary a folder for the application.properties (i.e. the files in Spring Boot containing configuration, e.g. URLs to web services, credentials).
While Java classes are composed on the class level, i.e. the fields within a Java class split up across multiple feature folders are merged, other artefacts like process models and application.properties files are merge by copying the artefacts into the same folder.
By default, Camunda Platform deploys all BPMN process files in the classpath resources folder.
To apply the configurations, we have configured Spring Boot in that all application.properties whose path in the classpath corresponds to a defined pattern are loaded.
After composition, the various artefacts (i.e. Java classes, process models, and application.properties) are arranged in the Maven standard structure so that the PAIS Product can be compiled natively with Maven.
At both start time and runtime, parameters may prevent certain implementations from being selected and executed.
In conditional statements in software programs, frequently parameters are used to decide whether or not a specific feature is executed based on the value of the parameter.
By providing values, parameters can be exploited as variability mechanism for feature binding at runtime.
Note that the use of parameters as variability mechanisms is well known in literature \citep{Gacek.2001,Jacobson.1997}.

Examples \ref{xmpl:poc:notify-craftsperson} and \ref{xmpl:poc:check-application} are used to demonstrate how static and dynamic binding of activity implementations works by registering them as plugins.
\begin{customExample} %
	\label{xmpl:poc:notify-craftsperson}
	(Feature Binding Time in PAIS Product Lines): Activity \textit{Notify craftsperson} in Figure \ref{example-process-poc} of business process \textit{special parking permit for craftsperson} is used to notify the applicant when his/her application was rejected.
	This activity is a variation point for which multiple implementations are available.
	The applicant may be notified via SMS, e-mail, or manually by a clerk.
	At compile time and start time, implementations may be selected by a developer.
	At runtime, the applicants can decide how they want to be notified from the available notification means that have been included by a developer.
\end{customExample}

\begin{figure}
	\centering
	\vfill
	\includegraphics[width=100mm]{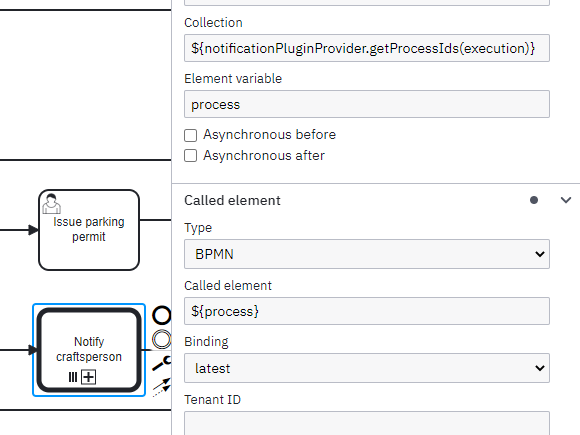}
	\caption{Camunda Configuration Activity \textit{Notify craftsperson}} \label{fig:camunda-modeler-config}
	\vfill
\end{figure}

Figure \ref{fig:camunda-modeler-config} shows the configuration of activity \textit{Notify craftsperson} for the Camunda Platform.
As aforementioned, activity implementations correspond to an implementation process model that embeds the implementing functionality (i.e. automated task or user task) and the corresponding code base, i.e. Java code to implement an automated task and HTML code for user tasks.
The Spring Service \textit{NotificationPluginProvider} provides the IDs of the implementation process models of all registered activity implementations.
Then, the Camunda Platform process engine iterates over the list of IDs using the iteration variable \textit{process} and calls the implementations process models.

The Spring Service \textit{NotificationPluginProvider} is depicted in Listing \ref{list:NotificationPluginProvider}.
It contains a field \textit{notificationPlugins} of type \textit{NotificationPluginRegistration}, which contains all registered implementations for notification activity.
When applying for the parking permit, the applicant may select one or more notification means from all registered ones.
The selection of the applicant is persisted in the process data structure.
When calling the method \textit{getProcessIds} (Listing \ref{list:NotificationPluginProvider}), all registered notification plugins are retrieved, iterated over, and checked whether the applicant has selected them. 
The method returns the IDs of the implementation process models that are registered and the user has selected.

\noindent\begin{minipage}{\textwidth}
\begin{lstlisting}[language=Java, basicstyle=\tiny, frame=single, numbers=left, caption=Notification Plugin Provider,
	label=list:NotificationPluginProvider]
@Service
public class NotificationPluginProvider {

  @Autowired
  private NotificationPluginRegistration notificationPlugins;

  public List<String> getProcessIds(DelegateExecution execution) {
    List<String> notificationWays = (List<String>) execution
	                             .getVariable("notificationWays");
    List<String> processIds = new ArrayList<String>();

    for (Entry<String, String> plugin : notificationPlugins
	             .getPluginProcessIdsByPluginId().entrySet()) {
      String pluginId = plugin.getKey();

      if (notificationWays.contains(pluginId)) {
        String processId = plugin.getValue();
        processIds.add(processId);
      }
    }
    return processIds;
  }
}
\end{lstlisting}
\end{minipage}

The Spring Service \textit{NotificationPluginRegistration} (Listing \ref{list:NotificationPluginRegistration}) has the field \textit{plugins} of type list, which picks up and injects all Spring Services that implement the interface \textit{NotificationPlugin}
illustrated in Listing \ref{list:notificationplugin}.

\noindent\begin{minipage}{\textwidth}
\begin{lstlisting}[language=Java, basicstyle=\tiny, frame=single, numbers=left, caption=Notification Plugin Registration,
	label=list:NotificationPluginRegistration]
@Service
public  class  NotificationPluginRegistration {
	
  @Value("${excludeNotification:}")
  private List<String> excludedPlugins = new ArrayList<String>();
	
  @Autowired
  private List<NotificationPlugin> plugins;

  public Map<String, String> getPluginLabelsById() {

    Map<String, String> pluginLabelsById = new HashMap<String, String>();
    for (NotificationPlugin plugin : plugins) {
      if (!excludedPlugins.contains(plugin.getId())) {
        pluginLabelsById.put(plugin.getId(), plugin.getLabel());
      }
    }
    return pluginLabelsById;
  }

  public Map<String, String> getPluginProcessIdsByPluginId() {

    Map<String, String> pluginProcessIdsByPluginId = new HashMap<>();
    for (NotificationPlugin plugin : plugins) {
      if (!excludedPlugins.contains(plugin.getId())) {
       pluginProcessIdsByPluginId.put(plugin.getId(), plugin.getProcessId());
      }
    }

    return pluginProcessIdsByPluginId;
  }
}
\end{lstlisting}
\end{minipage}

\noindent\begin{minipage}{\textwidth}
\begin{lstlisting}[language=Java, basicstyle=\tiny, frame=single, numbers=left, caption=Notification Plugin Interface,
	label=list:notificationplugin]
public  interface  NotificationPlugin {
	
  public String getId();	
  public String getProcessId();
  public String getLabel();
}
\end{lstlisting}
\end{minipage}

\begin{figure}
	\centering
	\vfill
	\includegraphics[width=100mm]{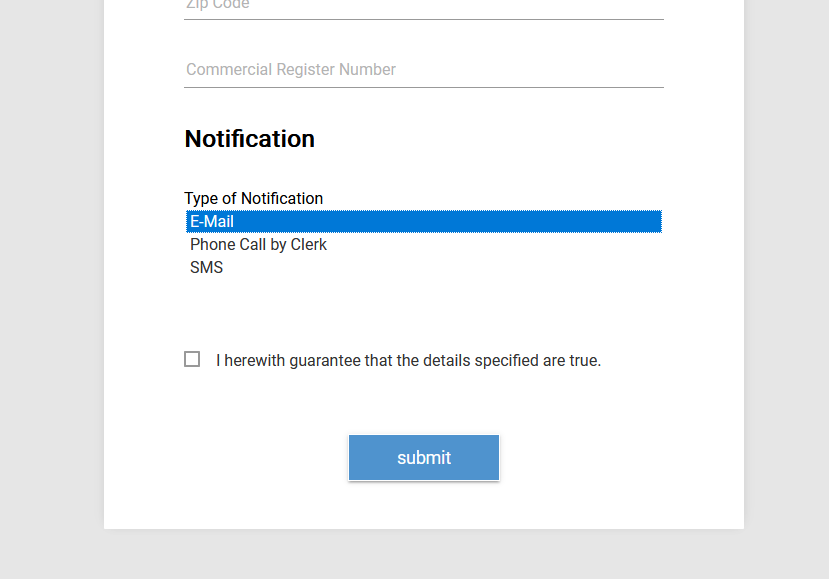}
	\caption{Extract of Application Form for Special Parking Permit} \label{fig:application-form}
	\vfill
\end{figure}

Every implementation of the notification activity needs to implement this interface in order to get registered as plugin.
The interface contains methods to get the ID of the plugin, the ID of the associated implementation process model, and the label of the plugin which is displayed in the application form (cf. Figure \ref{fig:application-form}).
Furthermore, the Spring Service \textit{NotificationPluginRegistration} has the field \textit{excludedPlugins}, which picks parameters up from command line during application start.
For example, the following command starts the Spring Boot application using \textit{Apache Maven} and specifies that the e-mail notification plugin shall be excluded and thereby not selectable in the application form.

\begin{lstlisting}
mvn spring-boot:run  -Dspring-boot.run.arguments=
    --excludeNotification=notification.mail
\end{lstlisting}

The Spring Service \textit{NotificationPluginRegistration} comprises two methods.
The first method iterates over the registered plugins, eliminates the excluded plugins, and collects and returns the labels of the remaining plugins.
This method is called via the Graphical User Interface (application form, cf. Figure \ref{fig:application-form}), which offers the applicant a selection of the available notification means.
The second method also iterates over the registered plugins, eliminates the excluded plugins, and collects and returns the IDs of the implementation process models of the remaining plugins.
The second method is called from the Spring Service \textit{NotificationPluginProvider}.

Figure \ref{fig:application-form} shows an extract of the application form for the special parking permit for craftspersons.
It can be observed that the applicant can choose the notification means.
E-mail, SMS, and notification by a clerk are available for selection in this extract, i.e. all plugins are included during compilation, are registered, and were not deselected during startup time.
This Graphical User Interfaces calls the Spring Service \textit{NotificationPluginRegistration}, gets the labels of the registered plugins and displays them.
When the applicant submits the application form, the IDs of the selected notification means are persisted in the process data structure such that only the selected notification means are used when notifying the applicant.

\begin{customExample} 
	\label{xmpl:poc:check-application}
	(Aggregation Codes for Special Parking Permit Process): Activity \textit{Check application data} in Figure \ref{example-process-poc}, which depicts business process \textit{special parking permit for craftsperson}, is used to check whether the parking permit shall be granted.
	This activity constitutes a variation point for which there are two available implementations.
	The application can be checked automatically by comparing the data from the application with the data from the commercial register.
	Besides, the application can be checked manually by a clerk.
	Both implementations write the result of the check to the process data structure.
	Consequently, if both implementations are selected the result data needs to be aggregated, i.e. aggregation code needs to be included that rejects the application if at least one implementation rejects the application.
\end{customExample}

\begin{figure}
	\centering
	\vfill
	\includegraphics[width=120mm]{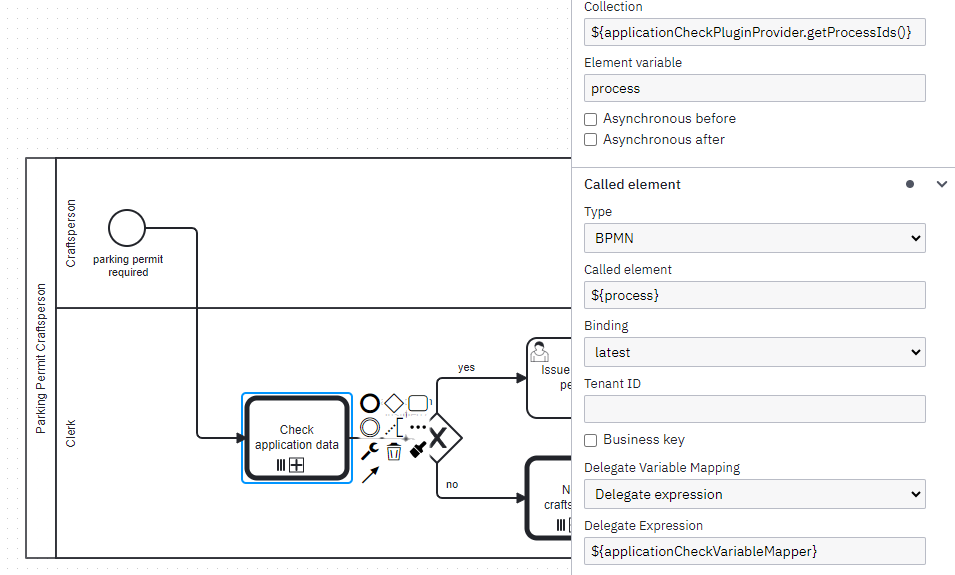}
	\caption{Camunda Configuration Activity \textit{Check application data}} \label{fig:camunda-modeler-config-check-application}
	\vfill
\end{figure}

When using subprocesses in Camunda, \textit{variable mappers} are implemented that specify what data and how the data from the subprocess is transferred to the parent process.
If only one implementation for activity \textit{Check application data} is selected during PAIS Product derivation, a simple variable mapper is included that passes field \textit{justified} from the subprocess to the parent process.
If both activity implementations are included, a variable mapper is included that collects the fields \textit{justified} from all activity implementations and evaluates them.
As soon as one activity implementation finds the application unjustified the variable mapper will pass \textit{unjustified} to the parent process.
A variable mapper is a Java class which is referenced in the process model by its Spring bean name and may be included using FeatureHouse.
In Figure \ref{fig:camunda-modeler-config-check-application}, the configuration for activity \textit{Check application data} is shown.
The IDs of the implementation process models that have been included during compile time are provided by a Spring bean and then used to call the implementation process models.
Furthermore, in section \textit{Delegate Variable Mapping} of the configuration the Spring bean \textit{applicationCheckVariableMapper} is referenced which acts as aggregation code.

From a technical perspective, the variable mapper (i.e. aggregation code) accepts the data fields \textit{justified} from each subprocesses and writes them to a list in the core process model.
Then, the variable mapper continuously iterates over the list and checks whether all values are true and therefore the application as a whole is considered justified.
Consequently, there are no race conditions as the order in which the subprocess complete does not matter.
In any case, the results of the subprocesses are added to a list and subsequently evaluated.
Subprocesses do not overwrite each others results.

Besides race conditions, during feature binding, there might occur other technical problems.
1) Data integrity might be impacted when executing multiple activity implementations and consequently writing process data concurrently (data corruption).
Furthermore, 2) activity implementations might be cancelled or unbound due to failures during execution (e.g. an activity implementation tries to call a web service that is not available).
These are typical problems not only when digitising business processes variants but also when digitising regular business processes for which most WfMSs provide mechanisms.
As mentioned before, Camunda Platform was not customised and call activities and subprocesses are used for feature binding at runtime.
Consequently, we are able to solely rely on Camunda Platform's error handling and resilience features.
In the following, it is shown how we use Camunda Platform to solve concurrency problems and errors during the execution of activity implementation.

1) In general, if multiple threads write concurrently data to a process, e.g. two activities complete at the same time, Camunda Platform throws an \textit{OptimisticLockingExceptions}, rolls back one of the transactions and retries\footnote{\url{https://docs.camunda.org/manual/7.22/user-guide/process-engine/transactions-in-processes/#handling-optimistic-locking-exceptions}}.
That is, one thread writes the data whereas the other retries writing the data a moment later.
When configured accordingly and applied to our approach, the activity implementations realised as subprocesses invoked by a call activity are each executed in an own thread, which results in concurrency.
Consequently, Camunda Platform manages the concurrent access to process data and hence data cannot be corrupted.

2) In general, if the execution of an activity fails regardless of the underlying error (e.g. a web service is not available), Camunda Platform uses a retry mechanism in order to continuously try executing the activity.
The repetition intervals and the maximum number of repetitions can be configured in Camunda Platform.
After the maximum number of repetitions has been reached, Camunda Platform raises an incident in the management interface, which is called \textit{Cockpit}.
Then, an administrator can decide how to proceed.
In most cases, when the error occurred due to a non-available web service, the execution can just be resumed.
However, the administrator has other options as well, e.g. change process data of the running process instance.
Note that during a rollback, data gathered since the last transaction boundary is lost.
Transaction boundaries and hence how far back to roll can explicitly be specified in business processes executed in Camunda Platform\footnote{\url{https://docs.camunda.org/manual/7.22/user-guide/process-engine/transactions-in-processes/#transaction-boundaries}}.
However, a failure during the execution of one subprocess of a multi-instance call activity does not impact the execution of others.
This retry mechanism applies to our usage of multi-instance call activity for feature binding as well\footnote{\url{https://docs.camunda.org/manual/7.22/user-guide/process-engine/the-job-executor/#use-a-custom-job-retry-configuration-for-multi-instance-activities}}.
Consequently, unless an administrator intentionally cancels a subprocess, features/activity implementations are not unbound during execution.

In summary, for every activity that constitutes a variation point there is a plugin mechanism.
Every implementation needs to implement the interface of the activity it wants to register with.
At compile time (i.e. static binding) activity implementations may be included or excluded using FeatureHouse.
The activity implementations consist of the implementation process model and the associated code base (Java code for automated tasks and html code for user tasks).
At start time, command line parameters, and, at runtime, user input may be used to deselect some activity implementations (dynamic binding).

\subsection{Tool Support}
\begin{figure}
	\centering
	\vfill
	\includegraphics[width=120mm]{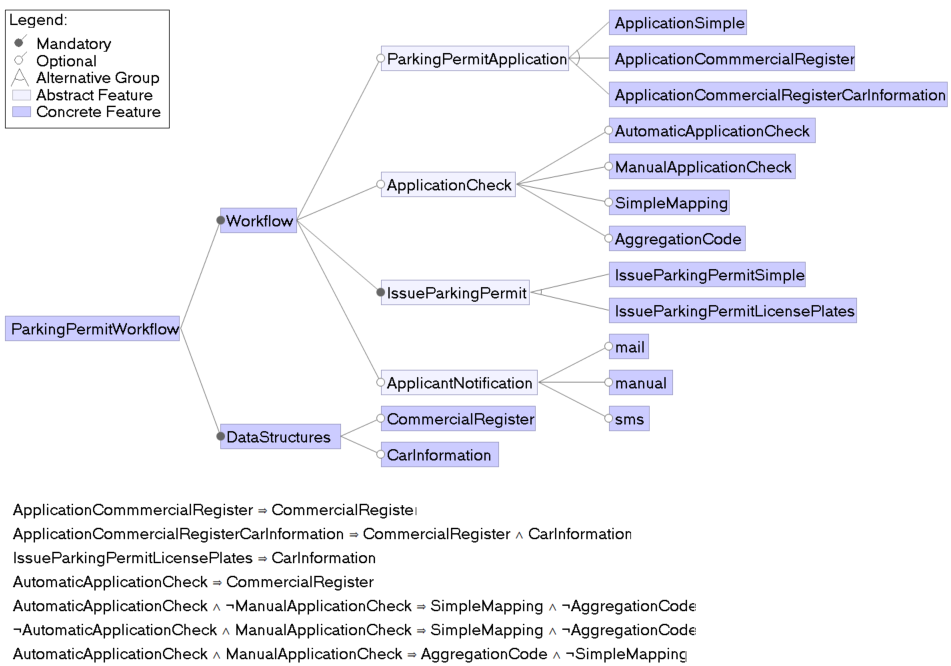}
	\caption{Feature Model in FeatureIDE} \label{fig:featureide-featuremodel}
	\vfill
\end{figure}

To support the phases of SPL Engineering with a tool, FeatureIDE is used in the proof-of-concept.
A feature model (cf. Figure \ref{fig:featureide-featuremodel}) visualises the variability of the PAIS Product Line during domain analysis.
Note that the feature model in Figure \ref{fig:featureide-featuremodel} is structured from left to right such that it fits on the page.
FeatureIDE captures the constraints in text form below the feature model as opposed to arrows used in other approaches.

\begin{figure}
		\centering
		\vfill
		\includegraphics[width=50mm]{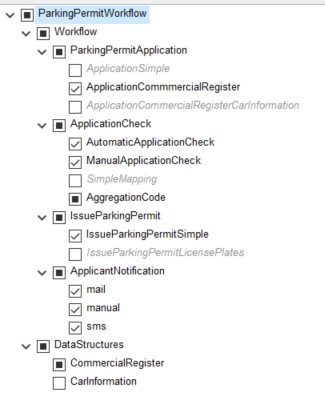}
		\caption{Feature Configuration in FeatureIDE} \label{fig:featureide-configuration}
		\vfill
\end{figure}

During requirements analysis, FeatureIDE supports the developer by providing a configuration editor (cf. Figure \ref{fig:featureide-configuration}).
In this configuration editor the developer can select the desired activity implementations that shall be included in the PAIS Product.
For implementing the variability of the PAIS Product Line, FeatureHouse was chosen, which is supported by FeatureIDE.
When deriving a PAIS Product, FeatureIDE uses the selection from the configuration editor as input (Figure \ref{fig:featureide-configuration}) to compose the variable activity implementations.

\section{Discussion}
\label{sec:discussion}
The presented approach has shown how to derive products from PAIS Product Lines with varying process data structures, which use multiple implementations for one activity, and that can bind features at compile time, start time, and runtime.
In the following, the approach is assessed in respect to whether it meets the requirements set out in Section \ref{sect:requirements}.
Furthermore, the proof-of-concept is tested to see whether it provides the listed properties.
Finally, limitations are disclosed.

\subsection{Requirements}
A PAIS Product Line corresponds to a set of common core artefacts from which PAIS Products may be derived. Hence, Requirement \textit{R1} is satisfied.
Multiple implementations can be selected for one activity. When the activity may write to the process data structure, aggregation code needs to be included in order to prevent the implementations from overwriting each others' data. Consequently, Requirement \textit{R2} is met.
Activity implementations may require adaptions to the process data structure.
When selecting a specific activity implementation, the required part of the data structure needs to be selected as well.
Therefore, Requirement \textit{R3} is satisfied.
Requirement \textit{R4} is met as a process feature model represents the variability of a PAIS Product Line with the corresponding constraints.
Unintended feature interaction can occur only when the activity implementations write to the process data structure. 
As aggregation code prevents implementations from overwriting each others' data Requirement \textit{R5} is satisfied.
At start time, activity implementations are selected per derived PAIS, which are not unbound while the PAIS is running.
Consequently, data cannot be lost.
At runtime, activity implementations are selected per process instance. Once they are selected, they cannot be unselected anymore.
Furthermore, activity implementations can be run independently and have no technical dependency among each other.
Consequently, activity implementations do not share system processes or data that might be corrupt. In addition, data does not get lost due to terminated activity implementations.
Hence, Requirement \textit{R6} is satisfied as well.  

\subsection{Properties}
The PAIS Product Line of the proof-of-concept uses FeatureHouse to include optional activity implementations.
When including an activity implementation, it registers as plugin with the corresponding activity.
Consequently, the proof-of-concept provides Properties \textit{P1} and \textit{P2}.
FeatureHouse is used to compose the process data structure as well. 
Hence, Property \textit{P3} is provided.
As a process feature model outlines the variability of the PAIS Product Line Property, \textit{P4} is provided.
Aggregation code is implemented as a Camunda variable mapper class, which is conditionally included using FeatureHouse.
Therefore, Property \textit{P5} is provided.
Static and dynamic feature binding is supported by using a mix of FeatureHouse and parameters.
Thus, Property \textit{P6} is provided.
Since FeatureIDE is used as tool support in the proof-of-concept Property \textit{P7} is provided.

\subsection{Scalability}
While the proof-of-concept is based on real-world requirements, we acknowledge the fact that there are business processes in German municipalities and industry with more activities and more implementation options.
Therefore, the scalability of the technical setup of the proof-of-concept is discussed in the following.

The phases of domain analysis, domain implementation, requirements analysis, software generation (i.e. phases of SPL Engineering), and runtime (i.e. execution of a derived PAIS) need to be distinguished. 
For the proof-of-concept, the modelling tool and the configuration editor of FeatureIDE were used for domain analysis and requirements analysis, respectively.
FeatureHouse, which is integrated into FeatureIDE, was used for domain implementation and software generation.
Camunda Platform is embedded as Java library in the proof-of-concept and executes the process models at runtime.
Consequently, in the following, it is shown how well FeatureIDE, FeatureHouse, and Camunda Platform scale.

FeatureIDE is used in industry, which is the result of a survey aimed at gaining insights about feature modelling and feature modelling tools, targeting exclusively practitioners with industrial experience \citep{Berger.01232013}.
In \citep{Lettner.93020151022015}, FeatureIDE was extended to be able to create hierarchical feature collections represented by feature models and define dependencies among features of the different collections.
Then, FeatureIDE was used to model two systems, which both have around 400 features of which some are mandatory, optional, and alternative.
Similarly, \cite{Linsbauer.2022} extended FeatureIDE to allow for partial configuration enabling splitting the product configuration among different groups of stakeholders so that the product is gradually configured.
This extension is now part of the official distribution of FeatureIDE, which we used for our proof-of-concept.
The authors evaluated their approach with five case studies of which one uses FeatureHouse as variability mechanism (i.e. the variability mechanism we use in our proof-of-concept) and contains 119 features.

\cite{Apel.2013c} have conducted 50 case studies in which they applied FeatureHouse to compose software artefacts to software systems.
The case studies include software systems written in various languages (e.g. Java) consisting of up to 99 features and approximately 64 thousand lines of code.
The authors conclude that FeatureHouse scales well with the number of features and lines of code.

After the artefacts have been composed using FeatureIDE and FeatureHouse, the proof-of-concept is natively compiled as Java application and Camunda Platform is used as an embedded Java library to execute the process models.
Camunda Platform has been benchmarked among other WfMSs \citep{Skouradaki.2016}.
The results show that the performance of Camunda Platform depends on the used workflow pattern.
However, generally speaking, Camunda Platform is able to execute in average more than thousand process instances per second when using a mix of workflow patterns.
The authors conclude that Camunda Platform performed well in comparison to the other investigated WfMSs.
Based on our own experience over the last six years working with Camunda Platform in industry and public sector, we can confirm that Camunda Platform is able to execute several tens of thousand processes concurrently.

In summary, FeatureIDE and FeatureHouse have been investigated in numerous case studies, in which they scaled well separately as well as in combination.
After the PAIS Product is derived (i.e. the artefacts are composed), the proof-of-concept is natively compiled as plain Java application, which includes Camunda Platform as WfMS.
Camunda Platform has proven suitable to execute process models in scaled settings.
As FeatureIDE and FeatureHouse are only used to compose the artefacts for compilation, they cannot influence the runtime.
Consequently, interferences between FeatureIDE, FeatureHouse on the one hand, and Camunda Platform on the other are not to be expected.
In conclusion, we are confident that the technical setup of our proof-of-concept as a whole scales well for business processes with lots of activities and various optional and alternative activity implementations.

\subsection{Limitations}
Using FeatureHouse, each feature corresponds to a folder, which contains the code of the feature.
As opposed to our previous work, the code of each feature (feature folder) cannot be compiled on its own.
When deriving a product, the code of the features is composed and after the composition the code is compiled.
Consequently, it is hard to develop the features independently.
Although each feature folder might constitute a repository under version control, it is difficult to reference a compatible version of this feature as it is no compiled artefact with a version. 

Furthermore, FeatureIDE currently lacks some convenience features.
Code completion is only available for selected features and there is no package view of the feature folders.
This is especially cumbersome for long package names as they are displayed as hierarchical folders instead of grouped packages.
\section{Evaluation}
\label{sect:evaluation}
In this section, we evaluate our approach with respect to 1) the relevance of digitisation in municipalities, 2) the suitability of existing standard PAISs for the special parking permit and other municipal business processes, and 3) whether our approach constitutes a solution to varying requirements of the special parking permit and other business processes among municipalities.

First, we evaluate our approach with expert interviews.
Then, threats to validity are discussed.

\subsection{Expert Interviews}
\label{sect:expert-interviews}

We conducted interviews with 15 experts most of whom are employed by German municipalities, two by a county, one was working as an external consultant for a municipality, and one is employed by a municipal umbrella organization, which is an association of municipalities that represents the interests of the municipalities to the state and federal government.
An overview of the experts is shown in Table \ref{tab:experts}.
The table contains the expert identifier (ID), the job title or department of the expert, the years of professional experience in the current position and in brackets the total years of professional experience.
The table also contains the number of inhabitants of the municipality or county the expert is employed by.

With the experts we interviewed, we cover all sizes of municipalities, from those with fewer inhabitants to a city with over a million of inhabitants.
To prevent conclusions from being drawn about the municipalities and thus the identity of the experts, we indicate the size of the municipalities in clusters (\textless50,000; \textgreater=50,000; \textgreater=75,000; \textgreater=100,000; \textgreater=300,000; \textgreater=1 million).
Due to the decreasing number of municipalities with larger populations, clusters representing higher inhabitant counts encompass a wider range of values.
We interviewed domain experts of the special parking permit for craftspersons, who are either a process participant, process owner or were responsible for the rollout of the corresponding digital process from a business perspective.
Furthermore, we interviewed experts working in IT or digitisation departments of municipalities most of which also have domain knowledge about the special parking permit as they were responsible for its rollout from a technical perspective.
Last but not least, we were referred to Expert K who is a domain expert in the area of road construction and who is currently responsible for the rollout of a PAIS allowing applicants to get a permit for road constructions.
As can be seen in the table, the interviewed experts are well-qualified to answer the questions, given that nearly all possess at least a few years of professional experience in their current positions, with some bringing up to several decades of total professional experience.
Experts A and F did not have specific knowledge about the special parking permit for craftspersons but rather could provide more general information about the digitisation approach of their municipality.
Expert K is a domain expert for road construction and Expert O is employed by a municipal umbrella organisation. 
Therefore, the answers of Experts A, F, K, and O regarding the special parking permit were not evaluated.
However, their general knowledge about digitisation of governmental services and business processes was valuable and their responses concerning this matter are included.
Consequently, we are confident to get valid feedback both regarding the different requirements of the special parking permit for craftspersons and varying requirements for municipal business process digitisation projects in general.

\begin{threeparttable}
	\centering
	\begin{tabular}{l|>{\raggedright\arraybackslash}p{7cm}|>{\raggedright\arraybackslash}p{1cm}|>{\raggedright\arraybackslash}p{4.0cm}}
		\hline
		\textbf{ID} & \textbf{Job Title/Department} & \textbf{YoE\tnote{1}} & \textbf{Inhabitants} \\
		\hline
		A & IT Department &  \multicolumn{1}{|c|}{1.5 (1.5)} & \textless  50,000 \\\hline
		B & IT Advisory Unit &  \multicolumn{1}{|c|}{4 (13)} & \textless  50,000 \\\hline
		C & IT and Organisation Department &  \multicolumn{1}{|c|}{3 (13)} & \textgreater= 50,000\\\hline
		D & Road Traffic Authority &  \multicolumn{1}{|c|}{1 (10)} & \textgreater= 50,000\\\hline
		E & Coordinator for Digitisation &  \multicolumn{1}{|c|}{3 (7)} & \textgreater= 50,000 \\\hline
		F & Chief Digital Officer &  \multicolumn{1}{|c|}{3 (12)} & \textgreater= 50,000 \\\hline
		G & Department for Traffic and Fines &  \multicolumn{1}{|c|}{7 (7.5)} & \textgreater= 75,000\\\hline
		H & Road Traffic Authority & \multicolumn{1}{|c|}{3 (3)} & \textgreater=100,000\\\hline
		I & Road Traffic Authority & \multicolumn{1}{|c|}{15 (36)} & \textgreater=100,000\\\hline
		J\tnote{2}  & Head of Traffic Authority & \multicolumn{1}{|c|}{11 (24)} & \textgreater=300,000\\\hline
		K\tnote{2} & Digital Construction Coordinator & \multicolumn{1}{|c|}{5 (30)} & \textgreater=300,000\\\hline
		L\tnote{3} & Economic Development & \multicolumn{1}{|c|}{8 (35)} & \textgreater=300,000\\\hline
		M\tnote{3} & Road Traffic Authority & \multicolumn{1}{|c|}{7 (8)} & \textgreater=300,000\\\hline
		N\tnote{4} & Solution Architect & \multicolumn{1}{|c|}{3.5 (10)} & \textgreater=1 million \\\hline
		O & Head of Digitisation & \multicolumn{1}{|c|}{3 (15)} & Municipal Umbrella Organisation \\\hline 
	\end{tabular}
	\begin{tablenotes}
		\item[1] Years of professional experience in the current position (total)
		\item[2] Both experts have the same employer
		\item[3] Both experts have the same employer, which is a county.
		\item[4] This expert is not employed by a municipality but was hired as an external expert for 3.5 years
	\end{tablenotes}
	\caption{Expert Panel Overview}
	\label{tab:experts}
\end{threeparttable}

With each expert, we conducted an online interview.
First, we introduced ourselves and disclosed the aim of our research to gain knowledge about varying requirements of the business process special parking permit for craftspersons and other municipal business processes.
Then, we asked the interview partner to answer questions about the importance of digitisation in municipalities, the process and data of the special parking permit as well as the used PAIS in the respective municipality and the degree of satisfaction with the latter.
Finally, we presented our prototype of the special parking permit and asked the interview partners whether the prototype could be used in their municipality/county only by applying the concepts presented in this paper (i.e. adapting process data structure, exchanging and using one or multiple activity implementation at different binding times).
Furthermore, we asked whether there are other business processes with varying requirements among municipalities and whether our approach can be applied to them as well.
The survey comprises 15 questions of which some are to be answered on a scale from 1 to 5 ("not at all" to "very").
We encouraged the interview partners to only answer the questions they felt fit to answer.
The interview questions are attached in \ref{appendix-questions}.
The notes taken during the interviews are available on GitHub\footnote{\url{https://github.com/hehnle/interviews-digitisation-in-german-municipalities}}. 

In the following, the results of the survey are presented.
The questions that could be answered on a scale (1-5) are evaluated in Table \ref{tab:experts-answers}.
On the left-hand side, an abbreviation of the question is shown.
The total number of answers to the question is shown in parentheses.
On the right-hand side, the distribution of the answers can be seen as a bar chart.
On the left-hand side of each bar, the total number of interview partners who answered \textit{not at all} or \textit{little} is shown, whereas the number on the right-hand side of the bar represents the interview partners who answered \textit{quite} or \textit{very}.

When being asked about the importance of digitisation in municipalities (cf. \ref{appendix-questions} Question 1), the interview partners uniformly answered that it was quite or very important.
Figure \ref{fig:reasons-importance-digi} shows the most frequently stated reasons for the importance, which include easier access for citizens to public services and faster processing of citizens' requests with less errors and therefore a facilitation of work for municipal employees.
However, Experts A and B point out that these benefits only come into effect if the business processes are digitised end-to-end without manual interruptions (i.e. data needs to be copied from one information system to another).
Expert N even is of the opinion that efficient and fast municipal processes are important for democracy with respect to state services supporting vulnerable citizens, i.e. financially disadvantaged citizens who are dependent on immediate state subsidies.

\begin{figure}
	\centering
	\begin{tikzpicture}
		\begin{axis}[
			symbolic x coords={easier access for citizens, 
				facilitation of work for municipal employees, 
				fewer errors, 
				faster process, 
			},
			ylabel={Number of Experts},
			xtick=data,
			bar width=15mm,
			axis line style={draw=none},
			nodes near coords, nodes near coords align={vertical},
			xticklabel style={ 
				align=center, 
				text width=4.3cm 
			},
			width=15.5cm,
			height=6cm,
			ymin=0
			]
			\addplot[ybar,fill=agree] coordinates {
				(easier access for citizens,   3) 
				(facilitation of work for municipal employees, 3) 
				(fewer errors, 3) 
				(faster process, 6) 
			};
		\end{axis}
	\end{tikzpicture}
	\caption{Reasons for the Importance of Digitisation Given by at Least 3 Experts.}
	\label{fig:reasons-importance-digi}
\end{figure}

As can be seen in Figure \ref{fig:kind-of-pais}, a majority of the municipalities surveyed (eight experts) use standard PAISs for the special parking permit process (cf. \ref{appendix-questions} Question 4).
Six of those eight experts state that they use an online portal provided by the state government containing a digital process for the special parking permit for craftspersons, which can be customised to some degree.
Expert H reports that they use a simple PDF file, which can be filled in and submitted via e-mail by a craftsperson, as they cannot use this online portal due to its limited customisation options (e.g. it is not possible to require the applicant to upload images of the car and the online platform requires the applicant to create an account).
Expert K affirms that they cannot use PAISs provided by state and federal government in their domain of road construction either as the degree to which the process is customisable is not sufficient (e.g. business process for the permit for telecommunication provider to roll out broadband).
Furthermore, four of the municipalities that use the state government portal additionally allow craftspersons to submit their application via e-mail.
Expert G even states that almost every applicant uses the PDF file instead of the online portal due to inconveniences when using the online portal (e.g. according to Expert B an account is required).
Contrarily, Experts N and J state that they use an in-house developed PAIS.
The functional scope of these information systems varies widely.
While the digital special parking permit process in the mentioned online portal only covers the application form and the decision step of a clerk, the information system provided by a private software vendor and used by Expert E and I provides a more holistic feature set including an automatic generation of the physical parking permit, which can be printed immediately, and an overview of all already issued parking permits.
The municipalities using the online portal provided by the state government need to copy the application data into a Word template or a separate information system to create and print a physical parking permit.
Generally speaking, Expert A and F state they use a variety of different PAISs for their business processes including PAISs provided by the state government, in-house developed PAISs and PDF files.
Furthermore, both municipalities are going to roll out a low-code platform in order to digitise their business processes.
Reasons for the low-code platform include the option to shape the digital process according to the needs of the municipality and only having to maintain one information system instead of individual systems for each business process.
Furthermore, for users (citizens and employees), it is easier to get used to just one information system.

\begin{figure}
	\centering
	\begin{tikzpicture}
		\pie[
		sum=auto,
		color={disagree, neutral, agree, sagree},
		text=legend 
		]{1/non-digitised, 2/In-house devloped PAIS, 4/Standard PAIS and non-digitised, 4/Standard PAIS}
		
	\end{tikzpicture}
	\caption{Software Adoption for the \textit{Special Parking Permit}}
	\label{fig:kind-of-pais}
\end{figure}

When invited to share the business process of applying, checking, and issuing the special parking permit for craftspersons (cf. \ref{appendix-questions} Question 2), the majority of interview partners described the process identically to how it is illustrated in Figure \ref{example-process-poc}.
Solely Expert C revealed that they require the craftsperson to pay the invoice before issuing the parking permit, whereas the other municipalities send the invoice to the craftspersons together with the parking permit.
Experts E, H and J state that they also intend to require payment before issuing the parking permit in the future.
While Experts C and H require an additional step (i.e. an activity in the business process) to check the payment, Experts E and J intend to introduce some kind of e-payment, which requires the applicant to pay before submitting the application form or before downloading the digitally issued parking permit, which does not necessarily equate to a new activity in the business process.
Expert N explains that they issue a provisional parking permit.
After having paid, the craftsperson receives the long-term parking permit.
As anticipated, the applicants need to provide different data for the parking permit depending on the municipality (cf. \ref{appendix-questions} Question 3).
Table \ref{tab:required-data} shows a comparison of the required data for applying for the special parking permit in the different municipalities.
While all experts confirm that the applicants need to provide the company name, contact information, and details about the cars, some of the municipalities require proof of the business indicating that the applicant is a craftsperson (e.g. by the chamber of industry and commerce).
Furthermore, some municipalities ask for images of the vehicles to verify that the parking permit serves the purpose and is not issued for the private car of the manager, whereas others do not ask for images or only ask for specific types of vehicles.
Besides, the maximum number of cars the parking permit will be valid for varies between the municipalities (up to four cars according to Expert I). 
Likewise, some municipalities need a specific reason (i.e. intended use/purpose) why the craftspersons need the parking permit to carry out their work.
Finally, Expert I states that the craftsperson can consent to be contacted by e-mail or phone.
Otherwise, they will be contacted via mail, e.g. to be informed of the rejection.

\begin{table}
	\centering
\begin{threeparttable}
	\def\arraystretch{1.3}
	{\setlength{\tabcolsep}{10pt} 
	\begin{tabular}{rccccccc}
		\textbf{Expert}                  & \textbf{B/D/L/M} & \textbf{C}  & \textbf{E}  & \textbf{G}  & \textbf{H}  & \textbf{I}  & \textbf{J} \\\hline
		Applicant's name                 & \checkmark       & \checkmark  & \checkmark  & \checkmark  & \checkmark  & \checkmark  & \checkmark \\
		Contact details                  & \checkmark       & \checkmark  & \checkmark  & \checkmark  & \checkmark  & \checkmark  & \checkmark      \\
		Vehicle Registration Certificate & \checkmark       &  \checkmark & \checkmark  & \checkmark  & \checkmark  & \checkmark  & \checkmark\\
		Image of Vehicle                &                  &  *          &             & **          & \checkmark  &             & \checkmark\\
		Intended use                     &                  &             & \checkmark  & \checkmark  & \checkmark  & \checkmark  & \checkmark\\    
		Proof of Business                & \checkmark       &             & \checkmark  &             & \checkmark  & \checkmark  & \checkmark\\ \hline
	\end{tabular}}
	\begin{tablenotes}
	\item[\checkmark] required
	\item[*] Required only for vans
	\item[**] Required only for atypical vehicle of craftspersons
	\item This table refers exclusively to the application form for craftspersons (i.e. other professions and features such as permit renewal/loss reporting are excluded). Furthermore, period/area of validity are excluded as there are too many differences.
\end{tablenotes}
\caption{Data Required to Apply for Special Parking Permit in the Municipality of the Respective Expert}
\label{tab:required-data}
\end{threeparttable}
\end{table}

A majority of experts think that their PAIS facilitates work little or not at all and that there is not at all or only little other improvement (cf. \ref{appendix-questions} Question 6 and 7 respectively).
Experts B, D, and J criticise the usability and lack of features of their PAIS, which is why using it does not facilitate their work at all.
According to Experts C and N a lack of technical interfaces requires clerks to copy and paste data from the application form to a Word template or another information system, respectively, to create and print the parking permit.
Consequently, for these experts, the digital process facilitates the work not at all or only moderately.
Experts E and I states that their PAIS facilitates the work to a very high degree due to technical interfaces with other software systems (e.g. automatic transfer of data)
Expert I highlights the option for the clerks to work from home.
As a result of the digital process, there are fewer errors, which is another improvement according to Expert I.
The answers of Experts H, L, and M are not included as Expert H does not use a PAIS yet and L and M have no comparison to the process before the digitisation.

When being asked whether the PAIS satisfies the experts' expectations regarding end-to-end process digitisation the answers vary greatly (cf. \ref{appendix-questions} Question 9).
While Experts G, L, M, and I are quite or very satisfied with their PAIS, Experts B, C, and N once again are little satisfied as their PAIS lacks technical interfaces to other information systems.
Expert B states that there are satisfying examples of end-to-end process digitisation provided by the state government, e.g. the certificate of residency can be requested online and be issued fully automatically without a clerk due to technical interfaces to the residents' registration office.
Experts D and J are not satisfied at all due to usability issues, lack of features, and also missing technical interfaces.

When being asked whether the PAIS was customised the experts answered very differently as well (cf. \ref{appendix-questions} Question 8).
Experts L and M are employed by a county, which customised the digital special parking permit provided by the state government and made it available to all municipalities that are part of the county.
Experts B and D are part of said county, use the process as is provided by the county and have, therefore, not customised the software at all.
Expert C uses the digital special parking permit of their county as well, which was quite customised.
Expert J is part of a regional association of municipalities, which developed a PAIS for the special parking permit on their own.
As the PAISs of Experts J and N are not the result of customised standard PAISs, their answers are not included in Table \ref{tab:experts-answers}.
Generally speaking, Expert A confirms that PAISs for other municipal processes should be customised as well.
However, frequently, the PAISs cannot be customised to the satisfaction of the municipality, which means that employees have to follow the process set out by the PAIS and cannot work in the way the process of the municipality envisages.

It can be observed that those municipalities that did customise the software to their needs are more satisfied than those that could not or did not.
For example, Expert I stated that they customised their software to a very high degree and are very satisfied with the result.
Expert G also states that they have customised the PAIS moderately and are quite satisfied.
Finally, Expert K is very satisfied with the PAIS in their domain of road construction in particular because of the customization options.
In contrast, Expert B and D did not customise their PAIS as they use the one provided by their county and are little or not all satisfied.

\begin{table}
	\begin{tabular}{l|l}
		Importance of Digitisation (14) & \likertpct{0}{0}{0}{4}{10}\\
		Facilitation of Work (8) & \likertpct{4}{0}{2}{0}{2}\\
		Other Improvement (8) & \likertpct{3}{1}{1}{3}{0}\\
		Satisfaction (10) & \likertpct{2}{3}{1}{3}{1}\\
		Customisation (7) & \likertpct{2}{1}{1}{2}{1}\\
		Features missing (9) & \likertpct{0}{3}{1}{4}{1}\\
	\end{tabular}
	\textcolor{sdisagree}{\rule{7pt}{7pt}} Not at all
	\textcolor{disagree}{\rule{7pt}{7pt}} Little
	\textcolor{neutral}{\rule{7pt}{7pt}} Moderately
	\textcolor{agree}{\rule{7pt}{7pt}} Quite
	\textcolor{sagree}{\rule{7pt}{7pt}} Very
	\caption{Experts' Answers Regarding their Digital Process \textit{Special Parking Permit for Craftspersons}}
	\label{tab:experts-answers}
\end{table}

Every expert misses some features at least a little, whereas a majority misses some features quite or very much (cf. \ref{appendix-questions} Question 10).
Figure \ref{fig:missing-features} illustrates the features that are missing according to the experts.
Experts C, D, J, and M would like to give the parking permit validity with a QR code in order to digitally issue the parking permit so that the parking permit does not have to be sent by mail and craftspersons can print it themselves.
The association of Expert J already issues parking permits digitally, however, without a QR code, which might result in illegal duplicates.
Experts L and I agree but highlight the difficulty of equipping all traffic wardens with devices for scanning the QR codes.
Experts C, D, G, and N would like an automatic transfer of the application data to the information system issuing the parking permit using a technical interface.
Expert B emphasises the advantages of a fully automated process which entails an automatic check of the application (i.e. technical interface with the commercial register) and an automated issuing of the parking permit, e.g. the number plate of the craftsperson is persisted in a database so that a physical parking permit is no longer required.
Then, the traffic warden can check the number plate during a traffic check.
Experts E and H confirm that, in the future, they also intend to issue fully automatically the parking permit either with a QR code or by persisting the number plate.
According to Expert E, there are already municipalities applying this approach.
While Expert F and O also underline the importance of fully automated processes, which include connecting to other government information systems and databases, they point out that these information systems and databases need to provide technical interfaces.
Other missing features are listed, such as the fact that tasks cannot be delegated to colleagues or that communication between colleagues via the information systems is not possible (according to Expert E).
However, these missing features are rather general requirements of an information system than requirements for the special parking permit and unrelated to process variability.
Finally, in the future, Experts E, H, and J intend to provide e-payment.

\begin{figure}
	\centering
	\begin{tikzpicture}
		\begin{axis}[
			symbolic x coords={technical interfaces, digitally issued parking permit, features unrelated to process variability, e-payment},
			ylabel={Number of Experts},
			xtick=data,
			bar width=15mm,
			axis line style={draw=none},
			nodes near coords, nodes near coords align={vertical},
			xticklabel style={ 
				align=center, 
				text width=4cm 
			},
			width=15cm,
			height=6cm
			]
			\addplot[ybar,fill=agree] coordinates {
				(technical interfaces, 6) 
				(digitally issued parking permit,  8) 
				(features unrelated to process variability,  2) 
				(e-payment, 3) 
			};
		\end{axis}
	\end{tikzpicture}
	\caption{Missing Features in PAIS for \textit{special parking permit} According to the Experts.}
	\label{fig:missing-features}
\end{figure}

After the experts provided information about their business process for the special parking permit, we presented our proof-of-concept and explained our concept to customise the process data, to exchange activity implementation and to select one or multiple activity implementations during different binding times.
All of the experts who answered the question (i.e. Experts B, C, D, E, G, H, I, J, L, M, N responded to \ref{appendix-questions} Question 11) confirmed that the proof-of-concept could be extended and customised to the needs of their municipality/county by only applying the presented concepts (i.e customising process data, selecting one to multiple activity implementation at different binding times).
Experts C and G once again highlighted that the last activity of the process needs to either automatically transfer the application data to an information system that is able to generate the physical parking permit or generate the parking permit itself.
Experts N and O agree and point out that connecting to other information systems is the biggest challenge.
Besides digitising a business process and its activities, Experts J and O remark that a municipal business process should be embedded in an online portal allowing communication between clerk and applicant and providing a login using an official account by the state or federal government.
Expert F affirms the applicability of our approach in general to municipal processes.
However, to enable a municipality to maintain a PAIS developed using this approach, the PAIS needs to be based on programming languages and technologies for which the municipality has skills.
Expert I points out that according to the security guidelines of their municipality, the application data completed in a publicly available form on the Internet must not be sent to the municipal intranet where a clerk will process the application.
We are aware of this security measure from one of our projects with a federal agency, where we changed the direction of the call to comply with the measure.
The application data is stored temporarily and retrieved from within the intranet at regular intervals.
This makes the firewall impenetrable from the Internet.
One challenge is set out by Experts C and N, who require an additional activity to check the payment before issuing the parking permit.
Currently, our approach does not support variation in the control-flow of a business process.
However, future work shall reveal how activity implementation variability and control-flow variability can be combined.

All of the experts who answered the question perceived our proof-of-concept as quite or very intuitive and easy to learn as can be seen in Table \ref{tab:experts-answers-poc} (cf. \ref{appendix-questions} Question 12).

\begin{table}
	\begin{tabular}{l|l}
		Intuitive and easy to learn (12) & \likertpct{0}{0}{0}{4}{8}\\
	\end{tabular}
	\textcolor{sdisagree}{\rule{7pt}{7pt}} Not at all
	\textcolor{disagree}{\rule{7pt}{7pt}} Little
	\textcolor{neutral}{\rule{7pt}{7pt}} Moderately
	\textcolor{agree}{\rule{7pt}{7pt}} Quite
	\textcolor{sagree}{\rule{7pt}{7pt}} Very
	\caption{Experts' Answers Regarding the Proof-of-Concept}
	\label{tab:experts-answers-poc}
\end{table}

When the experts were asked whether they know other business processes that vary among municipalities in similar ways like the special parking permit for craftspersons (cf. \ref{appendix-questions} Question 14), Experts B and E disclosed that almost all business processes vary among German municipalities.
While business processes such as the special parking permit for craftspersons are regulated by federal law and yet there are still differences, there are processes that only depend on municipal regulations and therefore there are even more differences.
These business processes include library cards and applications for municipal benefits (e.g. reduced costs for public swimming pools) for low-income families.
Expert C adds the business processes for poster permits (i.e. permission required to put up posters or advertisements in public spaces) and dog registration, for which they cannot use the PAIS provided by the state government as it cannot be customised to the degree to which it is necessary.
Expert F brings up the business process for travel requests for municipal employees as an example.
Expert K confirms the recurring problem in general that PAISs provided by state and federal government cannot be used due to their limited customizability.
Expert N even brings up a new perspective on varying requirements for business processes.
The responsibility of checking the parking permit for residents in the municipality he was working for varies. 
While there are districts in which special traffic wardens check the parking permits, in other districts the state police are responsible for checking the parking permit.
Originally, the municipality wanted to avoid issuing physical parking permits and instead only persist the number plate (like proposed by other municipalities for the parking permit for craftspersons).
However, the state police refused to install a mobile application on their smartphones to check the number plates.
Consequently, the parking permit needs to be issued physically for the entire municipality.
Using our approach, the activity implementation for issuing the parking permit could be selected depending on the district for which the parking permit shall be valid, i.e. for districts where the state police is responsible a physical parking permit is issued and for the remaining districts the number plate is registered in a database.
Experts A, B, C, E, F, H, N confirm that the presented approach can be applied to other municipal business processes as well whereas the other experts didn't feel confident to answer the question (cf. \ref{appendix-questions} Question 15).
As a final remark, Expert B argues if municipalities have to make compromises and still have to pay licence fees for a PAIS, they don't want to use it.

Questions 5 and 13 are not included in the evaluation as there were only a few responses and no conclusion could be drawn.

In conclusion, 1) the experts consider digitalisation in municipalities very important.
Furthermore, there are standard PAISs for the special parking permit and other municipal processes, which can be customised to some degree.
However, 2) the satisfaction of the municipalities regarding these PAISs varies widely.
Some municipalities even argue that they cannot use these standard PAISs at all or are not willing to pay for a PAIS that does not fully meet their expectations.
We acknowledge the fact that some of the reasons for the dissatisfaction have nothing to do with varying requirements for the business processes but rather the used information systems themselves, e.g. performance problems, missing features, or poor usability.
However, the experts repeatedly emphasised the necessity to customise the PAIS due to varying requirements including different process data, different ways of checking the application, a different way of issuing the parking permit, and a different notification means (i.e. selection of different activity implementation at different binding times).
Finally, the experts confirm that 3) our approach can be applied to the special parking permit for craftspersons and other municipal business processes to overcome the challenge of digitising business processes with varying requirements.
Besides variations of the activity implementations, some of the inspected business processes need variations in the control-flow.
Nevertheless, a majority of the experts confirm that our approach can be applied to their business processes as in these cases the control-flow is immutable and this majority can benefit from our approach.
Consequently, the feedback suggests that digitising business processes with varying requirements is not an individual case but rather a general issue and an important widespread challenge in German municipalities and our approach is suitable to mitigate this challenge.

\subsection{Threats to Validity}
We discuss the common four types of threats to validity according to \cite{Wohlin.2024} and how we tried to mitigate them: threats to internal validity, external validity, construct validity, and conclusion/reliability validity.

\textit{Threats to internal validity} might be false conclusions about causality.
For instance, we observed a relation between satisfaction with PAISs and the degree to which these PAISs can be customised to the needs of a municipality.
While we acknowledge other factors influencing satisfaction (e.g. performance, usability), experts have explicitly highlighted the relation between customisability and satisfaction.
In contrast to an anonymous questionnaire, the structured interview allows eliminating ambiguities in the conversation.
Furthermore, it was made sure that the experts answered the question regarding the applicability of the proof-of-concept in their municipality explicitly with respect to customisability and not usability.
The question of usability was asked separately.

An \textit{external threat to validity} is the lack of generalisability.
The question arises whether our approach can be generalised regarding technology and tools, special parking permits in municipalities, business processes in the domain of municipalities in general, and other domains.
Besides FeatureHouse and FeatureIDE, there are other tools for Software Product Line Engineering (some of which were presented in Section \ref{sec:fundamentals}), which are based on the same concepts.
Furthermore, Camunda Platform is a WfMS implementing common error handling and resilience features upon which we rely.
These features are implemented by other well-known WfMSs.
Especially, jBPM\footnote{\url{https://docs.jbpm.org/latest/jbpm-docs/html_single/}} and Activiti\footnote{\url{https://www.activiti.org/userguide/}} are similar as Activiti is based on the concepts of jBPM and Camunda is a fork of Activiti.
Consequently, while we have not implemented the proof-of-concept using other technologies, we are confident that our approach works independently of tools and technologies as we apply general concepts of Software Product Line Engineering.
A bias in the selection of the experts might prevent generalisation when the experts that are interviewed are too similar and consequently only one perspective is considered.
Therefore, we deliberately interviewed experts with different perspectives whose identities and backgrounds we could verify (e.g. their roles, years of experience, and type and size of their employer, i.e. municipality, county, or municipal umbrella organisation) including process participants, process owners, and experts for IT and digitisation in German municipalities.
Regarding the ability to generalise the challenge of digitising the special parking permit to a wider range of municipalities, the domain experts confirmed that this is a widespread challenge.
Besides the domain experts for the special parking permit, we also conducted interviews with IT Experts of municipalities, a county and a municipal umbrella organisation who have a good understanding in general of the requirements of municipal business processes that need to be digitised.
They confirmed that the challenge of digitising the special parking permit can be generalised to a variety of municipal business processes in Germany.
Although we interviewed a diverse group of experts with different perspectives, the group of 15 experts is rather small, which impacts generalisability.
While German municipalities each have the same responsibilities and offer the same services to the citizens in their geographical catchment area, German government institutions on a higher level (e.g. federal agencies) have individual and distinct responsibilities, i.e. their business processes differ completely.
Consequently, it is unlikely that our approach might be applied in German institutions at federal level.
However, based on our experience in process digitisation projects, we can confirm that the challenge of digitising business processes with implementation variability exists in the other domains (e.g. hospitals, insurance companies, HR business processes in the private sector) to which our approach can be applied.
For instance, in a current project at an insurance company, a business process is implemented that keeps a policyholder informed about a claim reported to the insurance company.
This business process differs between different branches of the insurance company in terms of connected third-party information systems and processed data.
As the branches have their own hardware, the digitised business process variants are deployed separately and shall only comprise the source code necessary for the corresponding branch.
This example shows that process variability in terms of differences in data structures and implementation of activities is not only found in the public sector and our approach is important for a number of organisations.
Systematic requirements analysis and case studies during future work shall provide evidence for the applicability of our approach to other domains.
 
\textit{Threats to construct validity} refer to the design of study or social threats.
The design is flawed when the measurement does not reflect the concept.
The wording of the questions may constitute a design threat as the questions may not be adequate to answer whether our approach represents a solution to the challenge of digitising business processes with varying requirements or the questions might be interpreted differently by the experts and the researchers.
To mitigate this threat, we formulated the questions based on prior-defined objectives (laid out at the beginning of Section \ref{sect:expert-interviews}) and took time to discuss and refine the wording internally considering knowledge and terminology we gained during digitisation projects in the public sector.
The objectives of the interview were not shared with the experts to prevent bias.
Besides, the experts might have had difficulties understanding the approach and the proof-of-concept in its entirety and the resulting implications (e.g. maintenance) and thus answer questions about its applicability due to the limited time in the interview.
While we have tried to mitigate the threat by describing our approach and proof-of-concept in detail, we are under the impression that the experts had a good understanding of the implications.
For instance, two experts pointed out that using our approach entails that the municipality needs to maintain another information system for which the necessary skills must be available.
Furthermore, as the experts could not answer the questions anonymously, they might have felt pressured to answer the questions according to what they believed was expected regarding their employer or the researchers, which constitutes a social threat.
To mitigate this threat, we assured the experts to use their answers anonymously.
There is a trade-off between anonymous surveys and interviews in which the experts need to reveal their identity.
As the experts revealed their identity to us we could verify the trustworthiness of the experts and make sure the answers are based on professional experience, which outweighs the social threat.
To further increase the validity of the answers, we encouraged the experts to skip questions when not being able to answer them confidently.

\textit{Conclusion validity} concerns the statistical power and significance of our observations.
As we use a qualitative evaluation, our data cannot be statistically evaluated and tested.
Therefore, \textit{reliability} is an issue.
We have taken utmost care in describing the procedure of the interviews, formulating the interview questions to avoid ambiguity, and outlining the backgrounds and composition of the experts as well as the size of the respective municipalities.
Furthermore, both the proof-of-concept and the interview questions are publicly available so that the experiment can be repeated.

\section{Related Work}
\label{sec:related-work}
In the following, related work is assessed.

\subsection{Single- and Multi-Model Approach}
\label{section:single-multi-model}

One way to deal with multiple variants of a business process is to include all variants in one process model by linking them with gateways \citep{Hallerbach.2010}.
However, process models containing multiple variants of the same business process may become very large and are thereby complex, unclear, and hard to maintain.

Besides combining variants into one process model, multiple process models for the various variants may be maintained \citep{Hallerbach.2010}.
Thus, similarities do not have to be investigated and the process variants can be independently modelled and evolve.
A challenge is to propagate a change to all process models, which is tedious and error-prone.

The aforementioned approaches deal with variability of process models rather than with variability on the activities' implementation level, on which our approach focuses.

\subsection{Customisable Process Models}
\label{section:configurable-model}
For managing lots of similar business processes, various approaches have emerged that allow configuring variability in process models by capturing commonalties and differences in one \textit{customisable process model} \citep{LaROSA.2017}.
Then, transformations can be applied to a customisable process model to derive a concrete process variant.
There are approaches that 1) allow removing elements from a customisable process model, 2) adding elements to a customisable process model, and 3) a combination of the aforementioned.
In the following, two approaches that are cited frequently are presented to illustrate the basic idea of customisable process models.

The approach of \textit{hiding and blocking} can be transferred to process models \citep{vanderAalst.2006}.
A reference process model contains all process variants from which concrete process models can be derived through configuration by hiding and blocking paths.
A process engineer may fill in a questionnaire to configure the reference process model \citep{LaRosa.2007}.
When an outgoing path of a process element is blocked during configuration, this path is not available any more at runtime, whereas a hidden path is skipped at runtime.

Instead of using a reference process model that contains all process variants,
\textit{PROVOP} \citep{AlenaHallerbach.2008,Hallerbach.2010,Hallerbach.2010b} uses a base process model to which various change patterns may be applied in order to configure concrete process models \citep{RinderleMa.2008}.
When modelling a base process, adjustment points have to be defined to indicate where adaptions may be applied in the model.
According to the PROVOP approach, possible adaptions include INSERT, DELETE and MOVE process fragments as well as MODIFY process element attributes \citep{Hallerbach.2010}.
In contrast to hiding and blocking, PROVOP allows also adding process behaviour at adjustment points during process configuration instead of just removing process elements (e.g. hiding and blocking).

Besides configuring the control flow of process models, other elements of a process model may be configurable. \cite{LaRosa.2011} propose an approach to individualise roles and objects modelled in a configurable process model.

For literature reviews on processes variability, the interested reader is referred to \citep{Ayora.2015,LaROSA.2017}.

The aforementioned approaches as well as the ones discussed in the literature reviews differ from our approach as they only provide means to configure elements of the process model like paths, roles, and objects, but neglect the configuration of activity implementations, i.e. these approaches do not take an software engineering perspective into account.

\subsection{Software Product Line Engineering for Variable Business Processes}
The concept of SPL Engineering has been applied to business processes in other works such as \cite{TaffoTiam.2015},\cite{Acher.2012}, \cite{S.Gimenes.2008}, \cite{Boffoli.2008}, and \cite{DanielSinnhofer.2015}.

In \citep{TaffoTiam.2015}, the authors introduce means to mark elements in a process model as variation points.
Different variants as well as their binding times may be specified as realisation for these variation points.
A proof-of-concept is built on top of the Eclipse plugin architecture.
However, the work focuses on the modelling aspect of variation points in process models rather than the implementations details of variation points.
The work lacks a detailed description of the variability mechanisms used to bind the variants and a description of the management of feature interactions.
In our approach, we present an in-depth approach on how to use existing variability mechanisms to bind features at compile time, start time, and runtime.
Furthermore, our approach describes how to handle feature interactions when applying concepts of SPL Engineering to variable business processes.

SPL Engineering is used by \cite{Acher.2012} as well to compose business processes.
The authors argue that the services called by a business process may be owned by a third-party vendor and may be highly configurable.
In order to compose a business process comprising services those services are each treated as an SPL whereas the business process itself is considered a multiple SPL.
The authors focus on aggregating the feature models of the SPLs, automatically checking the feature selection against the aggregated feature model, and determining which services are able to accept the output data of other services as input data.
Although the authors describe how to check whether or not the services are compatible (e.g. in regard to data structures), they fail to describe how the services are automatically composed, what technical mechanism shall be used to automatically bind and connect the services.
Furthermore, dynamic binding of the services and multiple binding for one activity are neglected by this work.
Besides, the considered business processes only include automated activities and do not contain user tasks.

Concepts of SPL Engineering like feature modelling have been applied to cross-organisation business processes by \cite{S.Gimenes.2008}. However, the approach is not holistic and does not describe how to systematically and automatically instantiate software products derived from the SPL.

In \citep{Boffoli.2008}, the customers' requirements and the corresponding parts of the business process that implement these requirements are mapped in a decision table, which can be used to configure and derive a business process from the product line.
However, the authors do not describe how the configuration works from a technical perspective.

SPL Engineering concepts were also applied to automatically generate process model variants.
Changes in process models can be automatically propagated to the derived process model variants \citep{DanielSinnhofer.2015}.
Our work distinguishes itself from the above one in that it focuses on the automatic generation of PAISs and not only on the automatic generation of process models.

\subsection{Software Process Lines}
SPL Engineering concepts have also been applied to software processes \citep{NogueiraTeixeira.2019}, which define the steps to develop software products (i.e. the process that describes how software development teams collaborate).
SPEM\footnote{\url{https://www.omg.org/spec/SPEM/2.0/PDF}} is a notation to define and illustrate software processes.
As SPEM models cannot be executed, there have been approaches to convert SPEM models to BPMN models in order to execute them, e.g. \cite{Dashbalbar.2017} manually translate a software process in the context of software testing modelled in SPEM to BPMN.

\cite{Agh.2021} have analysed Scrum\footnote{\url{https://scrumguides.org/docs/scrumguide/v2020/2020-Scrum-Guide-US.pdf}}, which is an agile approach commonly employed in software processes.
The authors identified the variation points in Scrum, which include roles, e.g. the Product Owner role may be exercised by an analyst or a project manager (both require domain knowledge of the target product).
Other variation points are the length of the development cycles (i.e. sprint length) and the technique for estimating the bearable workload for the next development cycles (i.e. sprint velocity).
The Scrum process is modelled using the notation SPEM.
Transformation rules are then automatically applied to the process model resulting in variants, whereby the resulting process models are manually adjusted.
In a case study, Scrum teams derived process variants according to their needs and adopted it in their daily work.
The authors argue their customisable Scrum software process contains best practices from which software development teams can select and adopt their best fit in order to address shortcomings in their current Scrum processes.
As the approach addresses a software process modelled in SPEM, the process model cannot be executed.

\cite{Aleixo.2010} introduced an approach that allows modelling a software process using the Unified Method Architecture (UMA), which is an evolution of SPEM 1.1.
Variability is represented as a feature model.
After deriving a process variant, the process variant can be transformed to a workflow specification that can be executed by the WfMS jBPM\footnote{\url{https://docs.jbpm.org/latest/jbpm-docs/html_single/}}.
All the activities of workflow specification are implemented as forms (i.e. user tasks).
While jBPM now uses BPMN as modelling notation to execute process models, at the time of the paper, the BPMN standard did not support native execution of process models \citep{Geiger.2018}.

\cite{Garcia.2015} pictured a software process using UMA for the development of service-oriented software products which can be tailored to the needs of a project.
The software process is first modelled incorporating variability.
Then concrete software processes may be derived by selecting the required elements at the variation points with respect to the needs of the project.
Finally, using a custom tool the software process model is converted to a BPMN model which is manually adjusted.
A WfMS can execute the converted BPMN model whose activities are all implemented by a standard form (i.e. user task).
While the converted process can be executed, the variability occurs solely in the process model.
There is no variability on the implementation level.
In addition, it is a very simple process consisting of only user tasks, which all comprise the same standard form.

The literature review conducted by \cite{NogueiraTeixeira.2019} revealed that most of the researchers applying SPL Engineering concepts to software processes use SPEM as a modelling notation for the control flow and only a few approaches (the two described above) allow executing processes.
In contrast to customisable process models, software process lines allow customising elements of the process (e.g. activities, roles) and rarely the control flow \citep{NogueiraTeixeira.2019}.
Finally, the authors point out that software process lines might benefit from further incorporating business process approaches, which allow executing the software processes.

In summary, the approach of software process lines focuses on variability in software processes whereas our approach addresses business processes in general, which has the following implications:
Software process lines include variation points that need to be implemented organizationally and which is not a technical challenge (e.g. staffing of a role).
Although some approaches allow converting process models to BPMN for execution, mostly the SPEM standard is used to model software processes, which lack the ability to be executed.
Consequently, software process lines only allow deriving process variants on an illustrative level.
In conclusion, our approach does not include software processes but business processes in general and allows automatically deriving entire information systems implementing business processes (i.e. PAIS Products) by solving variation points on an implementation level whereas software process lines focus on variability on the model level.
\section{Conclusions}
\label{sec:conclusion}
A PAIS Product Line comprises a common set of core artefacts including a core process model.
Activities of the core process model may be identified as variation points for which different implementations (e.g. user form, automated service) are developed.
PAIS Products may be derived from the PAIS Product Line by selecting implementations for the variation points.
Thereby, development efforts and costs are reduced when implementing similar PAIS Products sharing common features.

The presented work complemented the approach of PAIS Product Lines towards more flexibility.
A plugin mechanism was introduced that is based on FeatureHouse and parametrization.
Using this plugin mechanism, activity implementations may be bound at compile time, start time, and runtime.
Furthermore, FeatureHouse was used to compose the process data structure.
Different organisations may require varying process data structures for the same business process.
By composing the process data structure using FeatureHouse, the process data structure may be adapted to an organisation's specific needs during PAIS Product derivation.
In addition, the plugin mechanism enables the selection if multiple implementations for an activity.
It has been deduced that aggregation code is required when selecting multiple implementations for one activity that write to the same part of the process data structure to prevent the implementations from overwriting each others' data.
Aggregation code is additional code being included when multiple implementations for one activity are selected and that aggregates the output data of those implementations.
Finally, FeatureIDE is used as tool support for the phases domain analysis, domain implementation, requirements analysis, and product derivation.

We implemented the issuing of a special parking permit for craftspersons as a proof-of-concept, which is a common business process in German municipalities.
Then, we evaluated our approach and the proof-of-concept by conducting expert interviews which suggest that our proof-of-concept is applicable in many German municipalities and that digitising business processes with varying requirements for activity implementations is a widespread challenge in German municipalities.
Nevertheless, in order to test the applicability and practicability of our approach and tool chain, in future work, we want to carry out case studies in which we digitise further business process variants in order to asses the scalability and generalisability to other business processes and domains.

Furthermore, the presented approach has limitations, which will be addressed in future research.
It focuses on varying activity implementations and assumes a fixed and immutable control flow of the process model during PAIS Product derivation.
Although the expert interviews we conducted during our evaluation revealed that lots of German municipalities already can benefit from our approach, there are business processes whose control flow needs to be customised as well.
Future research shall reveal how the control flow may be tailored during PAIS Product derivation.
The approaches of base and reference process models might be combined and included into the approach of PAIS Product Lines. 
Besides, only activities may constitute a variation point, i.e only the implementation of activities may be selected and substituted.
Future research shall outline how events may constitute a variation point and how their implementation may be substituted.

\section*{Acknowledgement}
This research did not receive any specific grant from funding agencies in the public, commercial, or not-for-profit sectors.

\section*{Declaration of generative AI and AI-assisted technologies in the writing process}
During the preparation of this work the authors used Gemini in order to check spelling, grammar, and wording. After using this tool/service, the authors reviewed and edited the content as needed and take full responsibility for the content of the publication.

\bibliographystyle{elsarticle-harv}
\bibliography{bib}

\appendix
\section{Interview Questions}
\label{appendix-questions}

Introduction of interview participants and introduction to the topic.
\begin{enumerate}
	\item Do you consider the digitisation of business processes in municipalities important and do you consider it a facilitation of work?
	Answer options: 1-5 (1: not at all – 5: very much)
	\item Can you briefly explain the business process for issuing a special parking permit for craftspersons in your municipality?
	\item What information/documents are requested to apply for a special parking permit for craftspersons in your municipality?
	\begin{enumerate}
		\item[] Possible answers (others are allowed)
		\item Applicant's name
		\item Contact details
		\item Vehicle data
		\begin{enumerate}
			\item Registration certificate
			\item Photos of the vehicle, trunk, and tools
		\end{enumerate}
		\item Intended use for the special parking permit
		\item Business registration (e.g. confirmation by the Chamber of Commerce and Industry or extract from the commercial register)
	\end{enumerate}
	\item Do you use standard software? Do you use in-house developed software? 
	\begin{enumerate}
		\item[] Possible answers
		\item Standard software
		\item In-house developed software
		\item PDF/non-digitised
	\end{enumerate}
	\item Has your business process for issuing a special parking permit for craftspersons changed after digitisation? (Different steps/Different order/Different data) Answer options: 1-5 (1: not at all – 5: very much)
	\item Does the digital business process for the special parking permit for craftspersons facilitate work? If so, to what extent? Answer options: 1-5 (1: not at all – 5: very much)
	\item Have there been any other improvements, and if so, what are they? Answer options: 1-5 (1: not at all – 5: very much)
	\item If you use standard software, to what extent did it have to be customised (input fields/additional or other steps)? Answer options: 1-5 (1: not at all – 5: very much)
	\item To what extent does the information system you use meet your expectations for end-to-end process digitisation? Answer options: 1-5 (1: not at all – 5: very much)
	\item Are there any features you would like to have that your information system does not offer? And if so, which ones (e.g. a QR code to check the validity of the special parking permit)? Answer options: 1-5 (1: not at all – 5: very much)
\end{enumerate}
At this point in the interview, a demo of our proof-of-concept for the special parking permit for craftspersons is presented, including an explanation of the concepts for customising the business process.
\begin{enumerate}[start=11]
	\item Would the presented information system for the special parking permit for craftspersons be applicable in your municipality if it were tailored to your needs using only the presented concepts (customising data, exchanging activity implementations during various binding times)?
	\item Does the information system seem easy to learn and intuitive from an applicant's and municipal clerk's perspective? Answer options: 1-5 (1: not at all – 5: very much)
	\item Do you know of other municipalities that use the special parking permit for craftspersons in this or a different way?
	\item Do you know of other municipal business processes that differ between municipalities? If so, which ones and what are the differences?
	\item In your opinion, could the concept of the presented information system for the special parking permit for craftspersons be transferred to other municipal business processes (including reasons)?
\end{enumerate}

\end{document}